\documentclass[pra,showpacs,floatfix]{revtex4}
\usepackage{amsmath}
\usepackage{amssymb}
\usepackage{graphicx}
\usepackage{subfigure}

\begin{document}

\title{Exact solutions for periodic and solitary matter waves in nonlinear
lattices}
\author{C. H. Tsang$^{1}$, Boris A. Malomed$^{2}$, and K. W.
Chow$^{1}$} \affiliation{$^{1}$Department of Mechanical Engineering,
University of Hong Kong,
Pokfulam Road, Hong Kong\\
$^{2}$Department of Physical Electronics, School of Electrical
Engineering, Faculty of Engineering, Tel Aviv University, Tel Aviv
69978, Israel}
\begin{abstract}
We produce three vast classes of exact periodic and solitonic solutions to
the one-dimensional Gross-Pitaevskii equation (GPE) with the pseudopotential
in the form of a nonlinear lattice (NL), induced by a spatially periodic
modulation of the local nonlinearity. It is well known that NLs in
Bose-Einstein condensates (BECs) may be created by means of the
Feshbach-resonance technique. The model may also include linear potentials
with the same periodicity. The NL modulation function, the linear potential
(if any), and the corresponding exact solutions are expressed in terms of
the Jacobi's elliptic functions of three types, cn, dn, and sn, which give
rise to the three different classes of the solutions. The potentials and
associated solutions are parameterized by two free constants and an
additional sign parameter in the absence of the linear potential. In the
presence of the latter, the solution families feature two additional free
parameters. The families include both sign-constant and sign-changing NLs.
Density maxima of the solutions may coincide with either minima or maxima of
the periodic pseudopotential. The solutions reduce to solitons in the limit
of the infinite period. The stability of the solutions is tested via
systematic direct simulations of the GPE. As a result, stability regions are
identified for the periodic solutions and solitons. The periodic patterns of
cn type, and the respective limit-form solutions in the form of bright
solitons, may be stable both in the absence and presence of the linear
potential. On the contrary, the stability of the two other solution classes,
of the dn and sn types, is only possible with the linear potential.
\end{abstract}

\maketitle
\section{Introduction}

Due to the well-known fact that the Gross-Pitaevskii equation (GPE) provides
for an exceptionally accurate description of the dynamics of Bose--Einstein
condensates (BECs) in rarefied ultracold gases \cite{PitString03}, and the
related nonlinear Schr\"{o}dinger equation (NLSE) is an equally good model
of the light transmission in nonlinear media \cite{KA}, finding exact
solutions to these equations in nontrivial settings is a problem of great
significance \cite{general,general2,review}. A large class of exact periodic
solutions, which provide insight into the structure of one-dimensional (1D)
BEC patterns, was found in the framework of the GPE with specially designed
lattice potentials \cite{Carr,Carr2,Carr3,Carr4,Carr5,Carr6,Carr1}, that can
be created by means of available experimental techniques \cite{trapping}.
Exact solutions were also found for propagating density waves \cite{Kamch},
various localized states supported by particular spatiotemporal settings
\cite%
{localized,localized2,localized3,localized4,localized5,localized6,localized7, Kwok1,Kwok12,Kwok13,Kwok14}%
, stationary localized modes supported by the self-attractive nonlinearity
represented by sets of one or two delta-functions \cite{delta,delta2,Dong},
and for settings providing for the reduction of 3D states to the 1D form
\cite{Fedele}.

Another noteworthy class of exact solutions was recently reported in a model
combining linear and nonlinear lattices (NLs) \cite{Spain}. In the
experiment, linear periodic potentials can be induced by means of the
well-known optical-lattice techniques \cite{OL,OL2,OL3,Morsch}, while NLs
represent spatially periodic modulations of the nonlinearity coefficient,
which may be implemented by means of accordingly structured external fields
via the Feshbach-resonance effect. In fact, exact solutions reported in
Refs. \cite{delta,delta2,Dong} and \cite{Spain} are particular
representatives of a vast class of states supported by various patterns of
the spatial modulation of the nonlinearity (alias nonlinear \textit{%
pseudopotentials} \cite{pseudo}, in terms of the condensed-matter theory),
which were investigated in 1D \cite%
{nonlin,nonlin2,nonlin23,nonlin3,nonlin4,nonlin5,nonlin6,nonlin7,nonlin8,nonlin9,nonlin10}
and 2D \cite{2D,2D2,2D3,2D4,2D5} settings; mathematical aspects of
these
solutions were also studied in some detail \cite%
{math,math2,math3,math4}.

An issue of obvious interest is to find classes of \emph{exact}
periodic solutions supported by periodic pseudopotentials, and also
ultimate cases of such states with the infinite period, representing
matter-wave solitons. In addition, a challenging problem is to find
exact solutions that may be supported by combinations of a nonlinear
periodic pseudopotential and its linear counterpart of the
optical-lattice type. Of course, these states may be physically
meaningful only if they are stable. The objectives of the present
work is to report three classes of exact periodic solutions for the
GPE with a periodic pseudopotential, as well as for a more general
model including a linear periodic (lattice) potential, and to
analyze the stability of the periodic and solitonic solutions. We
find the form of the periodic functions accounting for the shape of
the NL pseudopotential and linear potential (if the latter is
included in the model), and, simultaneously, the respective exact
solutions by means of methods based on the Hirota bilinear
representation, which were elaborated, in the context of
``management" models for solitons, in Refs. \cite{Kwok1,Kwok12,Kwok13,Kwok14}%
, and for periodic (cnoidal) waves in various models with the bimodal
(two-component) structure and complex nonlinearities in Refs. \cite%
{Kwok2,Kwok21,Kwok22}. Making use of the basic Jacobi's elliptic functions,
\textrm{cn}, \textrm{sn}, and \textrm{dn}, we are able to find three
distinct families of nonsingular solutions, affiliated with each of these
functions; accordingly, the families are referred to as ones of the ``%
\textrm{cn}", ``\textrm{sn}", and ``\textrm{dn}" types. In the case
when only the NL pseudopotential is included, while the linear
potential is absent, each family is controlled by three free
parameters, two continuous and one which determines the sign of the
nonlinearity. If the linear-lattice potential is included too, each
family features \emph{four} free continuous parameters. In the limit
of the infinite period, the solutions of the \textrm{cn} and
\textrm{dn} types go over into bright solitons, while the respective
limit of the \textrm{sn}-type solutions is a dark soliton. The
stability of all these solutions, both periodic and solitary (except
for the dark solitons), is tested in this work by means of
systematic direct simulations of the perturbed evolution.

A noteworthy feature of the exact solutions of all the three types is that
they can be found in cases when the local nonlinearity periodically changes
its sign, thus forming NLs with the spatially periodic alternation of
self-focusing and defocusing layers. It is relevant to note that exact
solutions which were recently reported in a model combining nonlinear and
linear periodic potentials \cite{Spain} cannot be obtained in such a case,
as they are generated by a gauge transformation of the NLSE with constant
coefficients.

The normalized form and physical meaning of the GPE for the atomic wave
function, $\Psi (x,t)$, and of the NLSE for the local amplitude of the
electromagnetic wave in optics, are well known \cite{PitString03,KA} (in the
latter case, $t$ is the propagation distance of the beam launched into the
layered medium in the longitudinal direction, see Refs. \cite%
{optics,optics1,optics2,optics3,optics4}):
\begin{equation}
i\Psi _{t}+(1/2)\Psi _{xx}+g(x)\left\vert \Psi \right\vert ^{2}\Psi
-V(x)\Psi =0.  \label{GPE}
\end{equation}%
Here $g(x)$ and $V(x)$ represent the nonlinear and linear lattices,
respectively, with $g(x)>0$/$g(x)<0$ corresponding to the local
attraction/repulsion between atoms in the BEC. As usual, stationary
solutions to Eq. (\ref{GPE}) with chemical potential $\mu $ are sought for
as $\Psi \left( x,t\right) =\psi (x)\exp (-i\mu t)$, where real function $%
\psi (x)$ obeys equation
\begin{equation}
\mu \psi +(1/2)\psi ^{\prime \prime }+g(x)\psi ^{3}-V(x)\psi =0.  \label{psi}
\end{equation}%
The corresponding expression for the energy of the matter-wave configuration
is%
\begin{equation}
E=\int_{-\infty }^{+\infty }\left[ \frac{1}{2}\left( \psi ^{\prime }\right)
^{2}-\frac{1}{2}g(x)\psi ^{4}+V(x)\psi ^{2}\right] dx.  \label{E}
\end{equation}%
The analysis presented in this work is focused, first, on looking for
special forms of periodic functions $g(x)$ and $V(x)$, for which Eq. (\ref%
{psi}) admits exact solutions in terms of elliptic functions. Then, the
stability of the exact solutions is tested by means of simulations of the
perturbed evolution of those solutions within the framework of Eq. (\ref{GPE}%
).

The paper is organized as follows. In Section 2, we introduce the
class of exact solution of the \textrm{cn} type, first in the model
without the linear potential, and then in the its more general
version, including the potential. Results of tests of the stability
of this class of exact periodic solutions are reported in Section 3.
The \textrm{cn} patterns may be stable both without the linear
potential, and in the presence of the potential. Solutions of the
two other classes, of the \textrm{dn} and \textrm{sn} types,
together with the results of tests of their stability, are presented
in Sections 4 and 5, respectively. Both these species of the
patterns require the presence of the linear potential for their
stability. Section 6 reports results for bright solitons, which can
be obtained as ultimate-form solutions from families of the
\textrm{cn} and \textrm{dn }types. The bright solitons may be stable
in the absence and presence of the linear potential alike. Dark
solitons, which are limit-form solutions of the \textrm{sn} type,
are briefly considered too in Section 6, without the stability
analysis, which will be reported elsewhere. Conclusions drawn from
this work are formulated in Section 7.

\section{Exact solutions of the cn type}

\subsection{The model without the linear potential}

We first consider the case of the NL pseudopotential alone, with $V(x)=0$ in
Eq. (\ref{GPE}). Using the techniques elaborated in Refs. \cite%
{Kwok1,Kwok12,Kwok13,Kwok14,Kwok2,Kwok21,Kwok22}, which are based on
manipulations with the Jacobi's elliptic functions, it is possible to
identify general forms of the NL structural function, i.e., coefficient $%
g(x) $ in Eq. (\ref{psi}), which admit exact solutions to Eq. (\ref{psi})
with $V=0$, and the form of the solution, $\psi (x)$, itself. The first type
of the solutions is based on the elliptic cosine ($\mathrm{cn}$), the
corresponding expressions for $g(x)$ and $\psi (x)$ being%
\begin{equation}
g(x)=\frac{g_{0}+g_{1}\mathrm{cn}^{2}(rx)}{1+b\hspace{0.05in}\mathrm{cn}%
^{2}(rx)},  \label{g}
\end{equation}%
\begin{equation}
\psi (x)=A_{0}\frac{\mathrm{cn}(rx)}{\sqrt{1+b\hspace{0.05in}\mathrm{cn}%
^{2}(rx)}},  \label{solution}
\end{equation}%
The spatial periods of solution (\ref{solution}) and NL potential (\ref{g})
are, respectively,
\begin{equation}
L=4K(k)/r  \label{L}
\end{equation}%
and $L/2$, where $k$ is the modulus of $\mathrm{cn}$, which takes values $%
0<k\leq 1$, $K(k)$ is the complete elliptic integral of the first kind (the
case of $k=1$ corresponds to $L=\infty $, i.e., localized solutions).
Generally, expressions (\ref{g}) and (\ref{solution}) contain a set of four
free parameters, \textit{viz}., $g_{0}$, $r,$ $k,b$, while the remaining
coefficient (modulation depth), $g_{1}$, together with the amplitude and
chemical potential of the solution, are found to be
\begin{equation}
g_{1}=\frac{g_{0}b}{2}\frac{b(1+3b)+(1+b)(1-3b)k^{2}}{%
b(2+3b)-(1+b)(1+3b)k^{2}},  \label{g1}
\end{equation}%
\begin{equation}
A_{0}^{2}=g_{0}r^{2}[-b(2+3b)+(1+b)(1+3b)k^{2}],  \label{A}
\end{equation}%
\begin{equation}
\mu =\left( r^{2}/2\right) [1+3b-(2+3b)k^{2}].  \label{mu}
\end{equation}%
Using the obvious scaling invariance and assuming $g_{0}\neq 0$, we fix $%
\left\vert g_{0}\right\vert \equiv 1$ (if $g_{0}=0$, the above
solution becomes trivial). Throughout this paper, $g_{0}=\pm 1$ is
kept as the sign parameter. Thus, $k$ and $r$ determine the period
of the NL ($L$), $g_{0}$ is the overall sign of the nonlinearity,
and $b$ controls the depth of the spatial modulation of the periodic
pseudopotential. An obvious constraint on $b$, which may be both
positive and negative, is $b>-1$, which is necessary to avoid
singularities in Eqs. (\ref{g}) and (\ref{solution}). Further, the
remaining scaling invariance of Eqs. (\ref{GPE}) and (\ref{psi})
makes it possible to set $r\equiv 1$, without the loss of
generality, which is fixed below, unless the linear potential is
included. Thus, in the absence of the linear potential, the family
of the exact solutions of the \textrm{cn} type
depends on two free coefficients, $k$ and $b$, and the sign parameter, $%
g_{0} $.

Comparing expressions (\ref{g}) and (\ref{solution}), it is easy to see that
maxima of the density of the periodic solutions, $\psi ^{2}(x)$, which are
always collocated with points where $\mathrm{cn}^{2}(rx)=1$, coincide with
maxima or minima of $g(x)$--i.e., with, respectively, minima or maxima of
the NL pseudopotential--in cases when, severally, $g_{1}>g_{0}b$ or $%
g_{1}<g_{0}b$. Then, substituting expression (\ref{g1}) for $g_{1}$, we
conclude that, for $b<0$, the density maxima always coincide with maxima of $%
g(x)$, while for $b>0$ this is true under condition%
\begin{equation}
k^{2}<b/(1+b).  \label{max}
\end{equation}%
Otherwise, density \emph{minima} ($\psi =0$) are located at maxima of $g(x)$%
. In fact, taking into account condition $A_{0}^{2}>0$, as it follows from
Eq. (\ref{A}), one can conclude that condition (\ref{max}) may only hold in
the case of $g_{0}=-1$. Further, it is seen from Eq. (\ref{E}) that the
ground state, which realizes the minimum of the energy, \ must have density
maxima collocated with minima of the pseudopotential, i.e., maxima of $g(x)$
(of course, this condition is only necessary but not sufficient for the
exact solution to represent the ground state). It is relevant to mention
that various localized modes in the simplest model of the NL, represented
just by two spots at which the self-attractive nonlinearity is concentrated,
may be stable, even if they do not represent the ground state \cite%
{Dong}.

Lastly, we note that, as it follows from expression (\ref{g}), Eq. (\ref{psi}%
) degenerates into the ordinary GPE, with $g(x)\equiv \pm 1$, in two cases: $%
b=0$, or
\begin{equation}
g_{1}=g_{0}b.  \label{1b0}
\end{equation}%
In the former case, expressions (\ref{solution}) and (\ref{mu}) go over into
the usual cnoidal solution of the GPE with $g(x)\equiv +1$, i.e.,
\begin{equation}
\psi (x)=k~\mathrm{cn}(x),~\mu =(1/2)\left( 1-2k^{2}\right) ,  \label{b=0}
\end{equation}%
provided that $g_{0}=+1$ [for $g_{0}=-1$ and $b=0$, the solution does not
exist, as Eq. (\ref{A}) yields $A_{0}^{2}=-k^{2}$]. The other degeneration
condition, given by Eq. (\ref{1b0}), if combined with Eq. (\ref{g1}), yields
\begin{equation}
k^{2}=b/(1+b).  \label{b/(1+b)}
\end{equation}%
In this case, solution (\ref{solution}) is relevant for $g_{0}=-1$, being
\begin{equation}
\psi (x)=k~\mathrm{cn}(x)\mathrm{/dn}(x),
\label{limit}
\end{equation}%
[for $g_{0}=+1$, one obtains $A_{0}^{2}<0$ from Eq. (\ref{A})]. It is easy
to check that expression (\ref{limit}) is a straightforward exact solution
to Eq. (\ref{psi}) with $g(x)\equiv -1$ and $\mu =(1/2)\left( 1+k^{2}\right)
$, in agreement with Eq. (\ref{mu}).

\subsection{The model with the linear potential}

A more general class of exact periodic solutions of the \textrm{cn} type can
be found if the NL pseudopotential, introduced, as above, in the form of
Eqs. (\ref{g}) [but Eq. (\ref{g1}) should be dropped, see an explanation
below], is combined with the following linear-lattice potential:
\begin{equation}
V(x)=\frac{1+V_{0}\mathrm{cn}^{2}(rx)}{1+b~\mathrm{cn}^{2}(rx)},
\label{V(x)}
\end{equation}%
\begin{gather}
V_{0}=b+\frac{r^{2}}{2}[b(1+3b)+(1-3b)(1+b)k^{2}]  \notag \\
+\frac{3r^{2}(1+b)g_{1}}{2(g_{1}-g_{0}b)}[b-(1+b)k^{2}].  \label{V0}
\end{gather}%
The respective exact cnoidal solutions keeps the same general form as in Eq.
(\ref{solution}), but with the amplitude and chemical potential given by
\begin{equation}
A_{0}^{2}=\frac{3br^{2}(1+b)[b-(1+b)k^{2}]}{2\left( g_{1}-g_{0}b\right) }~,
\label{AV}
\end{equation}

\begin{equation}
\mu =1+(r^{2}/2)[(1+3b)-(2+3b)k^{2}].  \label{muV}
\end{equation}%
This solution family depends on \emph{four} continuous real parameters,
\textit{viz}., $g_{1},r,k,b$, and sign parameter $g_{0}$. Unlike the case of
$V=0,$ coefficient $r$ cannot be scaled out, unless $b=0$, and the value of $%
g_{1}$ is now another free parameter, rather than the one given by expression (%
\ref{g1}); in fact, Eq. (\ref{g1}) follows from Eq. (\ref{V0}) if one sets $%
V_{0}=b$, which reduces linear potential (\ref{V(x)}) to a trivial form, $%
V(x)\equiv 1$.

In the limit of $b\rightarrow 0$, one obtains, from the above formulas, $%
g(x)\equiv g_{0}$ and $V(x)=1+(1/2)\left( rk\right) ^{2}\mathrm{cn}^{2}(rx)$%
, the respective exact solution (\ref{solution}) being $\psi (x)=A_{0}%
\mathrm{cn}(rx)$ with $A_{0}^{2}=(3/2)g_{0}\left( rk\right) ^{2}$ and $\mu
=1-r^{2}\left( k^{2}-1/2\right) $, provided that $g_{0}=+1$. This is one of
exact solutions found in Refs. \cite{Carr,Carr2,Carr3,Carr4,Carr5,Carr6} for
Eq. (\ref{GPE}) with constant $g$.

\subsection{Further analysis of the exact solutions (without the linear
potential)}

The above solutions are meaningful provided that Eqs. (\ref{A}) and (\ref{AV}%
) yield $A_{0}^{2}>0$. Because the exact solution is defined in a vast
parameter space, several cases should be considered separately. In this
subsection, we do that for the solutions obtained in the absence of the
linear potential.

The first particular case corresponds to $g_{0}=+1$, $b>0$, see Eq. (\ref{g}%
). Then, as follows from Eq. (\ref{A}), condition $A_{0}^{2}>0$ requires
\begin{equation}
k^{2}>\frac{b\left( 2+3b\right) }{(1+b)(1+3b)}.  \label{k^2}
\end{equation}%
For all $b>0$, Eq. (\ref{k^2}) is compatible with constraint $k<1$. On the
other hand, Eq. (\ref{k^2}) is incompatible with condition (\ref{max}),
hence this subfamily of the exact solutions cannot represent the ground
state (and, in fact, it cannot be stable, as shown below).

As said above, an interesting possibility is to find exact solutions in the
\emph{sign-changing} NL, which makes the model drastically different from
the usual NLSE. As follows from Eqs. (\ref{g}) and (\ref{g1}), the
NL-modulation coefficient, $g(x)$, periodically changes its sign for $%
g_{1}<-1$, i.e.,
\begin{equation}
\left[ \frac{b(1+3b)+(1+b)(1-3b)k^{2}}{b(2+3b)-(1+b)(1+3b)k^{2}}\right]
\frac{b}{2}<-1.  \label{sign}
\end{equation}%
In view of underlying condition (\ref{k^2}), the denominator on the
left-hand side of Eq. (\ref{sign}) is negative, hence this inequality,
together with Eq. (\ref{k^2}), identifies a parameter region in which the
exact solutions are available for the sign-changing NL:%
\begin{equation}
\frac{b\left( 2+3b\right) }{(1+b)(1+3b)}<k^{2}<\frac{b(4+3b)}{(1+b)(2+3b)}.
\label{interval}
\end{equation}%
It is easy to check that, for $b>0$, interval (\ref{interval}) always
exists, and, as a whole, it lies within the region of $0<k<1$.

At small values of $b$, region (\ref{interval}) is squeezed into a narrow
strip, $2b-5b^{2}<k^{2}<2b-7b^{2}/2$, where NL-modulation function (\ref{g})
takes the form of $g(x)\approx 1+g_{1}\cos ^{2}x$ [recall that, while
considering Eq. (\ref{GPE}) with $V=0,$ we set $r=1$ in all expressions
following from Eqs. (\ref{g}) and (\ref{solution})], and exact solutions (%
\ref{solution}), (\ref{A}) become quasi-harmonic ones, with a small
amplitude, $\psi (x)\approx b\sqrt{-3/\left( 2g_{1}\right) }\cos x$, the
corresponding chemical potential being $\mu \approx 1/2$. In this asymptotic
form of the solution, $g_{1}$ may be treated as an arbitrary parameter $\sim
\mathcal{O}(1)$, the solution existing \emph{solely} for $g_{1}<0$, which
includes the case of the sign-changing NL for $g_{1}<-1$. It is evident that
this explicit asymptotic solution has its density maxima coinciding with
minima of $g(x)$ (\emph{maxima} of the pseudopotential), which are located
at $x=\pi n$, $n=0,\pm 1,\pm 2,...$ .

In the opposite limit, $b\gg 1$, existence region (\ref{interval}) is narrow
too, $1-2/\left( 3b\right) <k^{2}<1-1/\left( 3b\right) $%
. The respective modulation profile, $g(x)$, and exact solution (\ref%
{solution}) reduce to chains of ``flat-top" solitons.

The case of $g_{0}=+1$ and $-1<b<0$ in the nonlinearity-modulation function (%
\ref{g}) can be considered too (recall the solutions with $b<0$ may
represent the ground state). The inspection of Eq. (\ref{A}) shows
that exact solution (\ref{solution}) exists for all $k^{2}<1$ at
$0<-b<1/2$, and, at $1/2<-b < 2/3$, it exists in interval
\begin{equation}
k^{2}<\frac{|b|\left( 2-3|b|\right) }{\left( 3|b|-1\right) \left(
1-|b|\right) },  \label{inverse}
\end{equation}%
cf. Eq. (\ref{k^2}), there being no solutions at $-1<b<-2/3$. Further
considerations of Eqs. (\ref{g}) and (\ref{g1}) demonstrate that the NL
cannot change its sign in this case, always corresponding to the local
self-attraction.

The other generic subfamily of the exact solutions is obtained for $g_{0}=-1$%
. First, considering this case with $b>0$ in Eq. (\ref{g}), condition $%
A_{0}^{2}>0$ imposes a constraint on the elliptic modulus,
\begin{equation}
k^{2}<\frac{b(2+3b)}{(1+b)(1+3b)},  \label{another}
\end{equation}%
which always complies with $k<1$. Further straightforward considerations
demonstrate that constraint (\ref{another}) does not admit sign-changing
NLs, i.e., in the case of $g_{0}=-1$ and $b>0$, the exact solutions pertain
to the spatially modulated nonlinearity which remains self-repulsive at all $%
x$.

Another possibility is to take $g_{0}=-1$ and $-1<b<0$ in Eq. (\ref{g}). A
simple algebra demonstrates that the corresponding condition $A_{0}^{2}>0$
[see Eq. (\ref{A})], combined with $k<1$, is satisfied with any $k$ for $%
2/3<-b<1$, and is never satisfied for $0\leq -b<1/2$. In the intermediate
case of $1/2<-b<2/3$, the exact solutions exist in interval
\begin{equation}
k^{2}>\frac{|b|\left( 2-3|b|\right) }{\left( 3|b|-1\right) \left(
1-|b|\right) },  \label{shown}
\end{equation}%
cf. Eq. (\ref{inverse}). The consideration of expression (\ref{g1})
demonstrates that condition $g_{1}>1$ automatically holds in the present
case, hence NL modulation function (\ref{g}) is always of the sign-changing
type. Thus the exact solutions in this case may, in principle, represent the
ground state of the BEC with nonzero mean density, loaded into the
sign-changing NL. In fact, it will be demonstrated below that solely in this
case the stationary periodic solutions of the \textrm{cn} type may be \emph{%
stable} (without the addition of the linear potential).

It is relevant to stress that the existence of the exact \textrm{cn}-type
solutions with $g_{0}=-1$ is a nontrivial finding in the following sense:
according to Eqs. (\ref{solution}), (\ref{g1}), (\ref{A}), and (\ref{mu}),
in the limit of $b=0$ the same type of the solution, but taken with $%
g_{0}=+1 $, goes over into the commonly known exact \emph{unstable} cnoidal
solution for the GPE with $g(x)\equiv +1$ (the uniform self-attraction),
given by expressions (\ref{b=0}). Thus, the $\mathrm{cn}$ solutions for $%
g_{0}=+1$ may be regarded as a continuation of this simple unstable
solution, which helps to explain the fact that they are always unstable too,
as shown below. On the other hand, in the case of $g_{0}=-1$ the existence
range of the exact solutions is separated, as demonstrated above, by
interval $0\leq -b<1/2$ from $b=0$ [in particular, at $b=0$ and $g_{0}=-1$,
Eq. (\ref{A}) yields $A_{0}^{2}=-k^{2}$], i.e., this subfamily of the
\textrm{cn} solutions has no counterpart among solutions of the GPE with the
constant nonlinearity coefficient, and it does not originate from that
limit--in particular, it may be stable for this reason.

\section{The stability of the \textrm{cn}-type solutions}

The test of stability of the exact solutions is a crucially important part
of the analysis. In this work, we do not aim to compute stability
eigenvalues for small perturbations around stationary solutions. Instead, we
focus on systematic direct simulations of perturbed solutions. Although less
rigorous mathematically, this approach is closer to the description of the
physical situations, where finite random perturbations are always present.

The numerical integration was performed in the spatial domain with the
length corresponding to $100$ periods of the structure, see Eq. (\ref{L}),
and periodic boundary conditions. The simulations were run up to time $%
t=5000 $, in terms of the present notation. For most patterns, which
have the spatial period $L\leq 5$ (see below), this implies $t >
100$ characteristic dispersion times, estimated as
$T_{\mathrm{disp}}\sim L^{2}$. In fact, in all cases the
instability, if any, manifests itself at a much shorter time scale,
$t<500$. The simulations were extended to $t=5000$ to make sure that
solutions identified as a stable ones do not develop a delayed
instability (still longer simulations, carried out for selected
cases, completely corroborate the stability). Initial random
perturbations, at the $1\%$ amplitude level, were always sufficient
to unambiguously categorize stable and unstable solutions (the
application of stronger perturbations did not change the results).

\subsection{The solutions without the linear potential}

\begin{figure}[tbp]
\center\subfigure[]{\includegraphics [width=7.0cm]{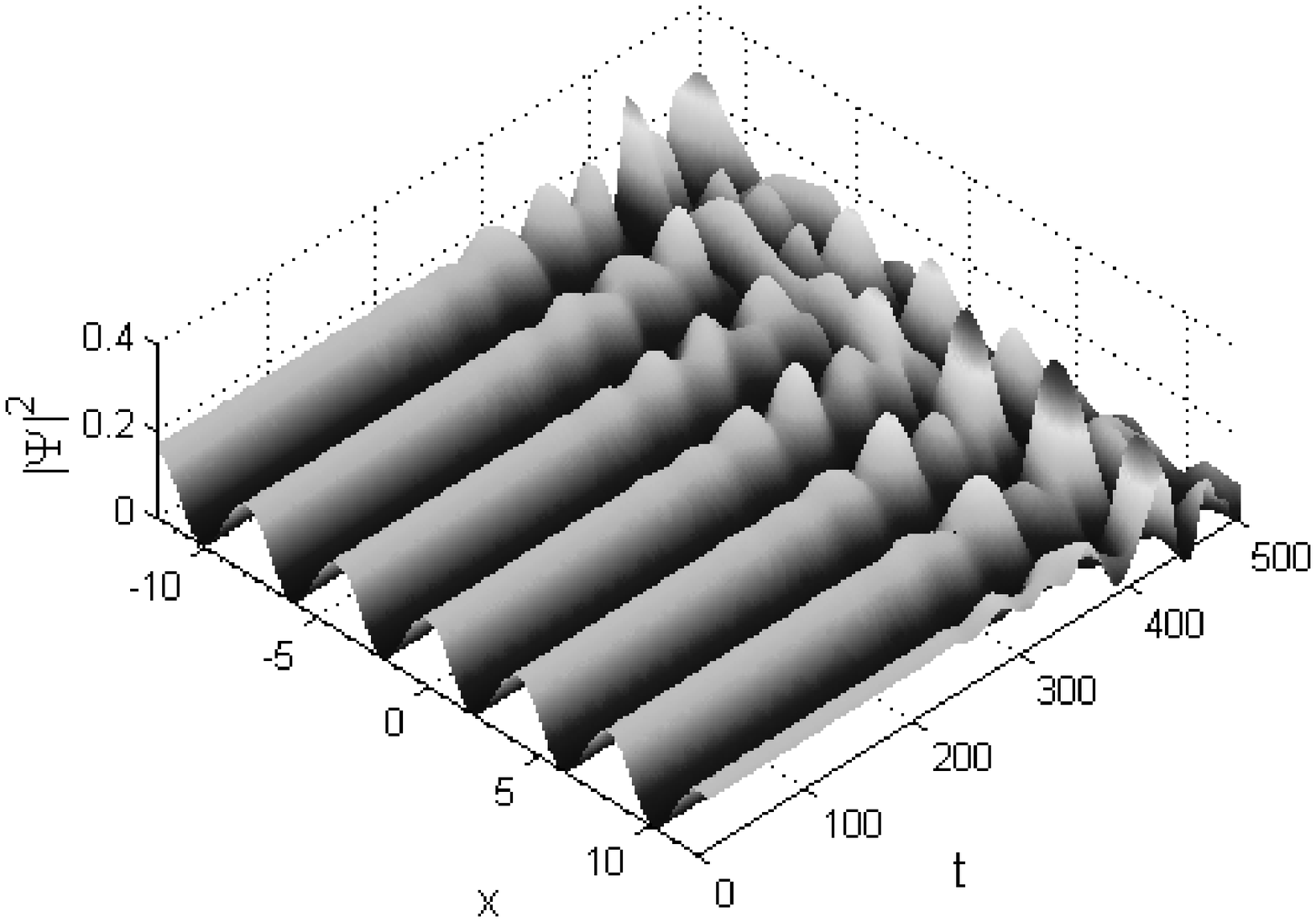}}%
\subfigure[]{\includegraphics [width=5.0cm]{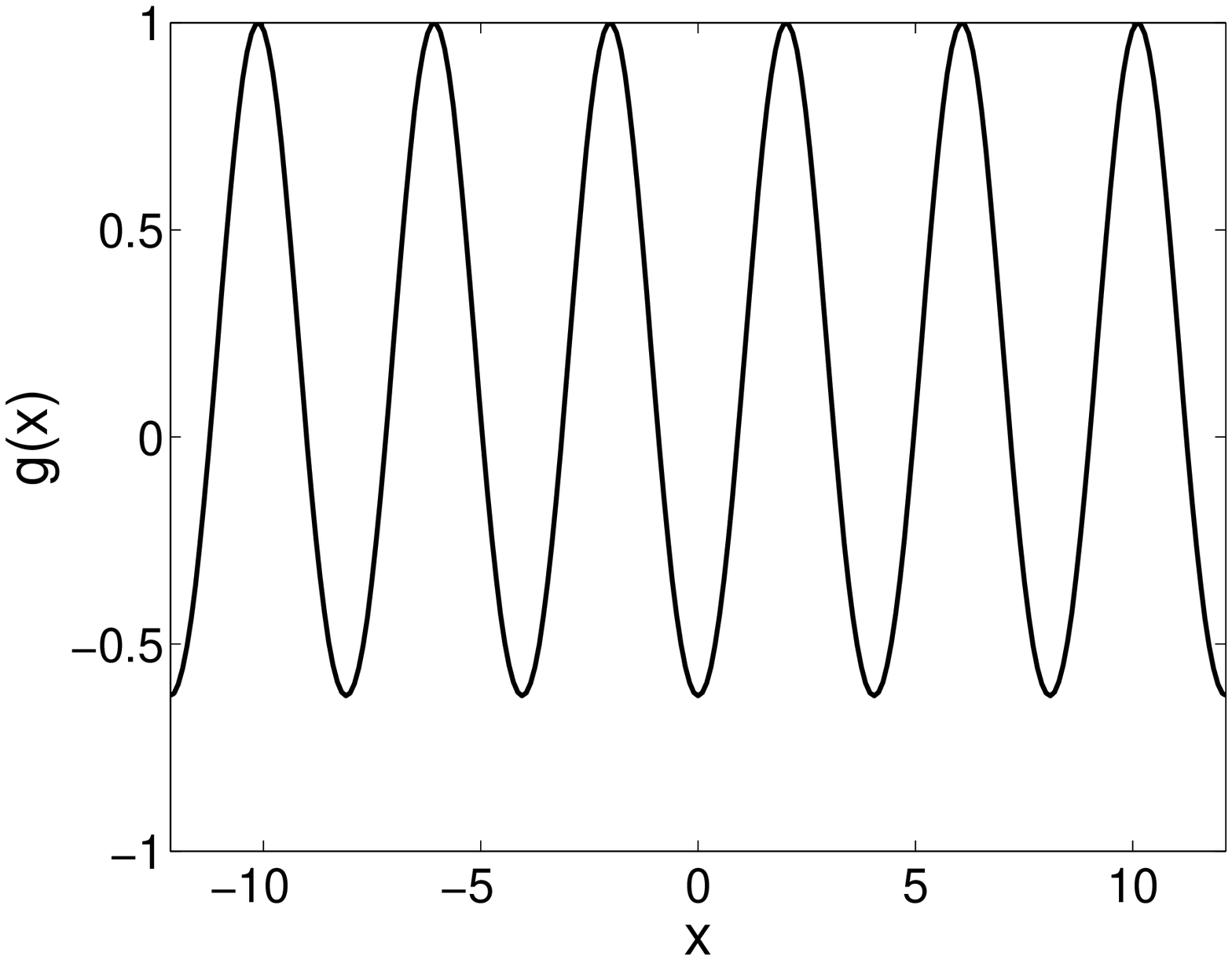}}
\caption{(a) The spatiotemporal evolution of the density of a
numerical solution to Eq. (\protect\ref{GPE}) demonstrates the onset
of the instability of the exact periodic solution of the \textrm{cn}
type, induced by a weak initial random perturbation (at the level of
$1\%$ of the amplitude of the stationary pattern). (b) The
corresponding profile of the periodic nonlinearity-modulation
function $g(x)$. The respective values of parameters in Eqs.
(\protect\ref{g}), (\protect\ref{solution}), and (\protect \ref{L})
are $g_{0}=+1$, $b=1$, $r=1$, and $k=0.82$. This exact solution was
obtained in the absence of the linear potential, with $g_{1}=-2.25$
given by Eq. (\protect\ref{g1}). } \label{fig1}
\end{figure}

We start the presentation of the stability results by considering the
\textrm{cn}-type solutions in the absence of the linear potential, $V(x)=0$
in Eqs. (\ref{GPE}) and (\ref{psi}). First of all, in all cases with $%
g_{0}=+1$ and $b>0$ (which suggests that the nonlinearity is
self-attractive, on the average), the simulations reveal that the \textrm{cn}%
-type solutions are unstable, see a typical example in Fig. \ref{fig1}. In
accordance with results of the above analysis, density maxima of the
unperturbed solution in Fig. \ref{fig1}(a) coincide with minima of $g(x)$ in
Fig. \ref{fig1}(b), i.e., maxima of the respective pseudopotential, which
readily explains the instability. Note that, although only six spatial
periods of the stationary solution are shown in the figure, it is actually a
representative fragment of the picture which was generated in the domain
covering $100$ periods, as said above. The same pertains to other figures
displayed below.

A generic feature observed in Fig. \ref{fig1} is that the instability sets
in after evolution time $t\leq 10T_{\mathrm{disp}}$. The instability onset
may be delayed for the unperturbed solutions with a small amplitude, which
is an obvious manifestation of the nonlinear nature of the instability.

In the same case of $g_{0}=+1$ but with $b<0$, unambiguously stable
stationary solutions were not found either. The difference of this case from
the one with $b>0$ is that the respective NL function $g(x)$ is a
sign-constant one, i.e., the nonlinearity is locally attractive everywhere
(which makes the modulational instability of stationary periodic solutions
quite plausible).

\begin{figure}[tbp]
\center\subfigure[]{\includegraphics [width=7.0cm]{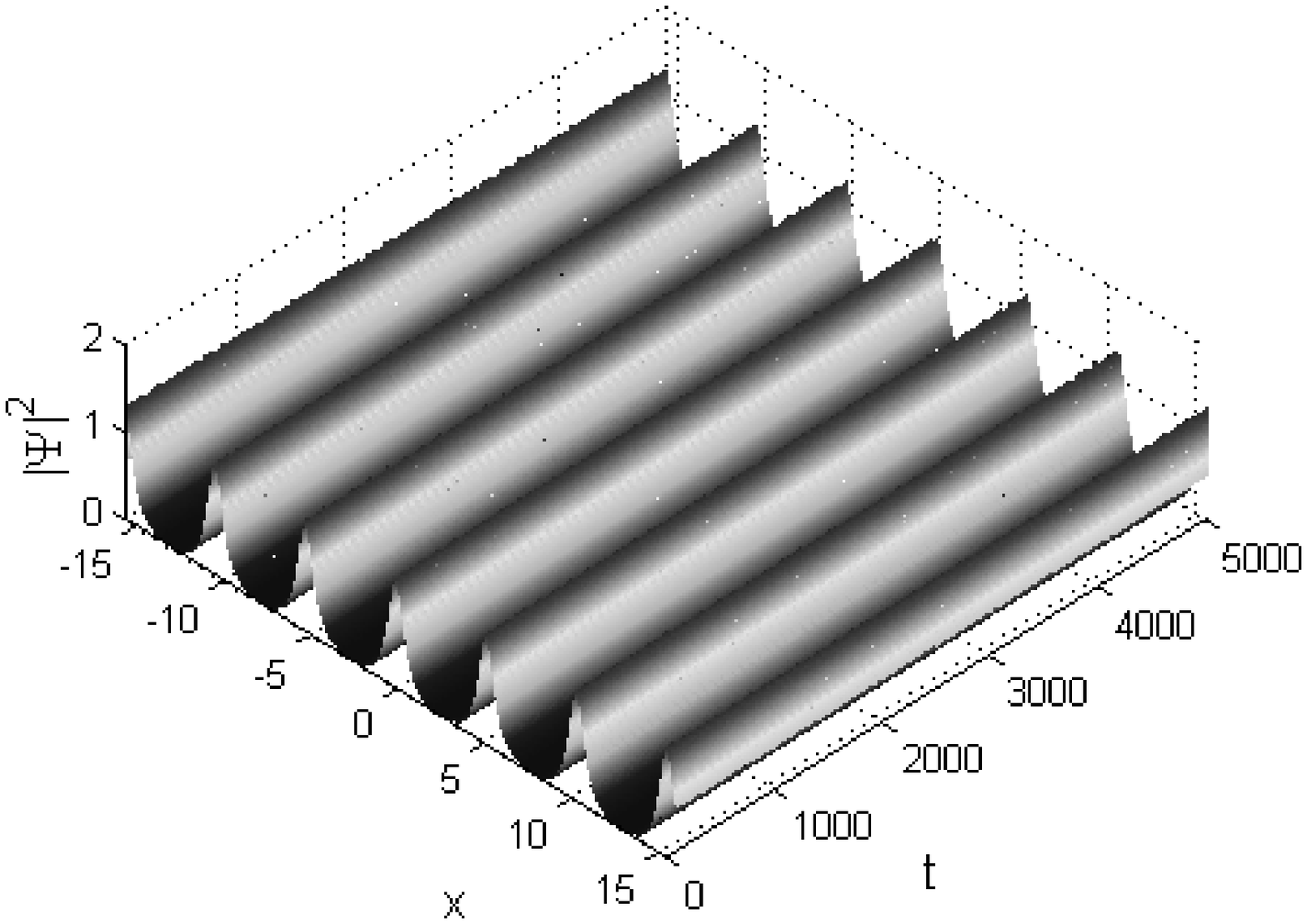}}%
\subfigure[]{\includegraphics [width=5.0cm]{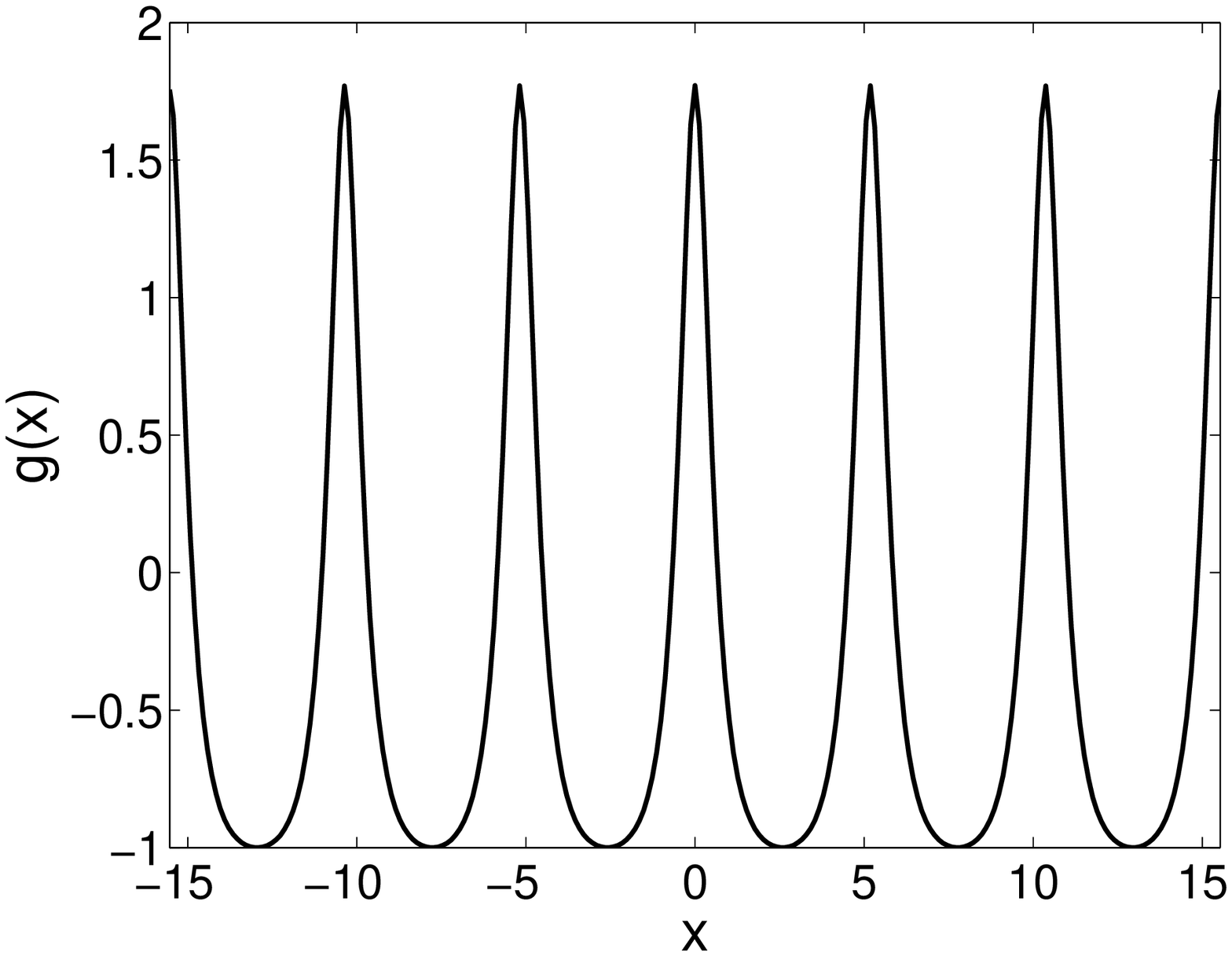}}
\caption{The meaning of panels (a) and (b) is the same as in Fig.
\protect \ref{fig1}, but in this case an example of a
\emph{marginally} stable
spatially periodic solution of the \textrm{cn} type is shown, for $g_{0}=-1$%
, $b=-0.7$, $r=1$, $k=0.95$, and $g_{1}=1.53$.}
\label{fig2}
\end{figure}

Proceeding to the case of $g_{0}=-1$, i.e., the model with the
nonlinearity which tends to be self-repulsive on the average, no
stable solutions have been found for $b>0$ in Eq. (\ref{g}).
However, stable solutions are possible if $g_{0}=-1$ is combined
with $b<0$ and appropriate values of $k$, which should be relatively
close to $k=1$ [recall that the respective exact solutions exist in
the interval of $1/2<-b<-1$, and they all correspond to a
sign-changing function $g(x)$]. An example of a stable pattern is
shown in Fig. \ref{fig2}. In fact, this figure displays a
\textit{marginally stable} solution, in the sense that it is never
destroyed by the originally imposed random perturbation, but the
disturbance does not dissipate either, remaining trapped in the
solution. Note that this example shows density maxima collocated
with maxima of $g(x)$, i.e., minima of the corresponding NL
pseudopotential, which facilitates the stabilization of the pattern,
and gives it a chance to be a ground state.

\begin{figure}[tbp]
\center\subfigure[]{\includegraphics [width=7.0cm]{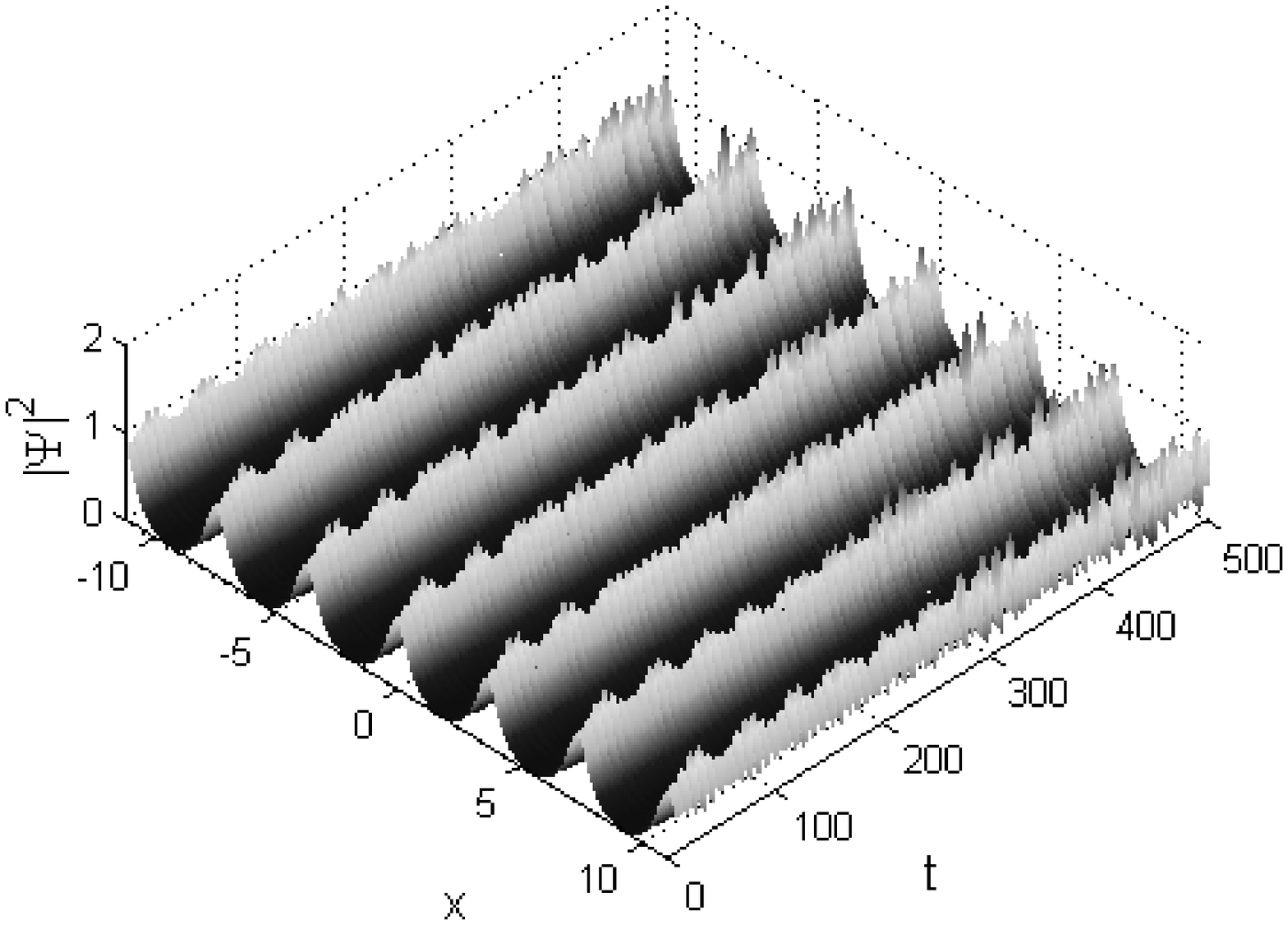}}%
\subfigure[]{\includegraphics [width=5.0cm]{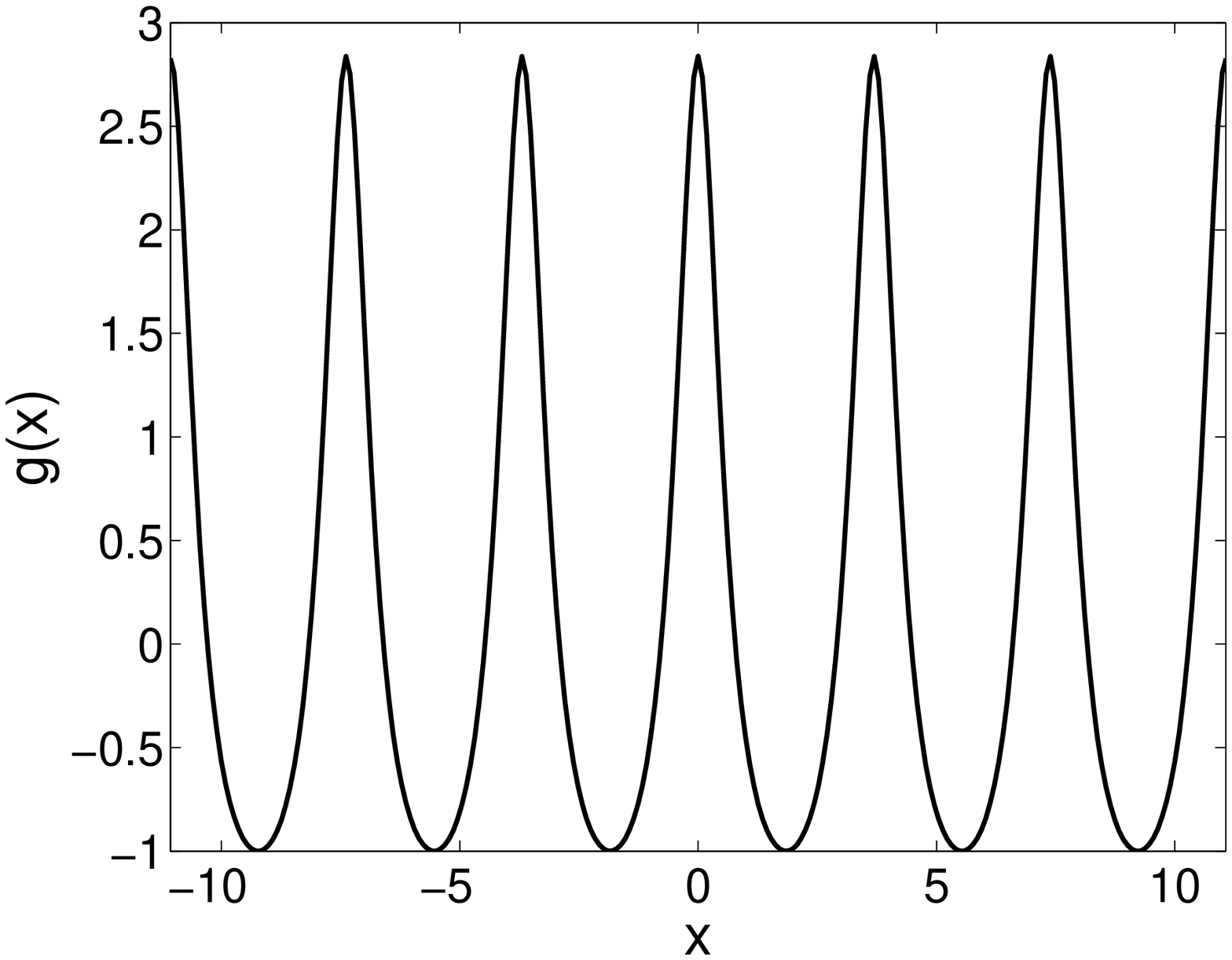}}
\caption{The same as in Fig. \protect\ref{fig2}, but for $k=0.70$
and the corresponding value of $g_{1}=1.85$, as given by Eq.
(\protect\ref{g1}). In this case, the \textrm{cn}-type solution is
unstable.} \label{fig3}
\end{figure}

Nevertheless, in the case of $g_{0}=-1$ and $b<0$ the \textrm{cn} solutions
may be unstable if the elliptic modulus is taken too far from $k=1$. For
instance, Fig. \ref{fig3} displays an example of a quick onset of the
instability, found at the same value of $b=-0.7$ as in Fig. \ref{fig2}, but
with $k=0.95$ replaced by $k=0.70$.

\begin{figure}[tbp]
\center\includegraphics [width=8cm]{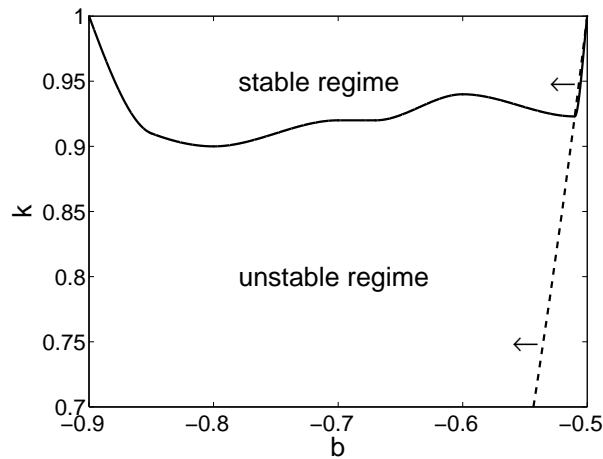} \caption{The region
of the stability of the \textrm{cn}-type periodic stationary
solutions, as found for $g_{0}=-1$ without the linear potential, in
the plane of parameters $b$ and $k$, which are defined in Eq.
(\protect \ref{g}). The dashed oblique line with arrows indicates
the border of the
area where the solutions exist, as per Eq. (\protect\ref{shown}), for $%
1/2<-b<2/3$.}
\label{fig4}
\end{figure}

Collecting data of the systematic numerical simulations for the presently
considered subfamily of the exact \textrm{cn}-type solutions, with $g_{0}=-1$%
, $1/2<-b<1$ and various values of $k$, makes it possible to identify a
\emph{stability area} in the plane of $\left( b,k\right) $, as shown in Fig. %
\ref{fig4}. The stability boundary in this figure was drawn by interpolating
results obtained from the direct simulations for
\begin{equation*}
b=-0.9,~-0.85,~-0.8,~-0.75,~-0.7,~-0.67,~-0.65,~-0.6,~-0.55,~-0.51.
\end{equation*}

As said above, the stability is limited to relatively long-wave patterns,
with $k$ sufficiently close to $1$, \textit{viz}., $k\geq 0.9$, the
respective values of period (\ref{L}) being $L\geq 10.12$. In accordance
with the above analysis, the existence region of the solutions is limited to
$1/2<-b<1$. It may be relevant to notice that the stability boundary in Fig. %
\ref{fig4} is not monotonous, which may be understood as a consequence of
the fact that the solution depends on $b$ in a complex manner, as seen from
Eqs. (\ref{g1}) and (\ref{A}).

\subsection{The \textrm{cn}-type solutions in the presence of the linear
potential}

\begin{figure}[tbp]
\center\subfigure[]{\includegraphics [width=7.0cm]{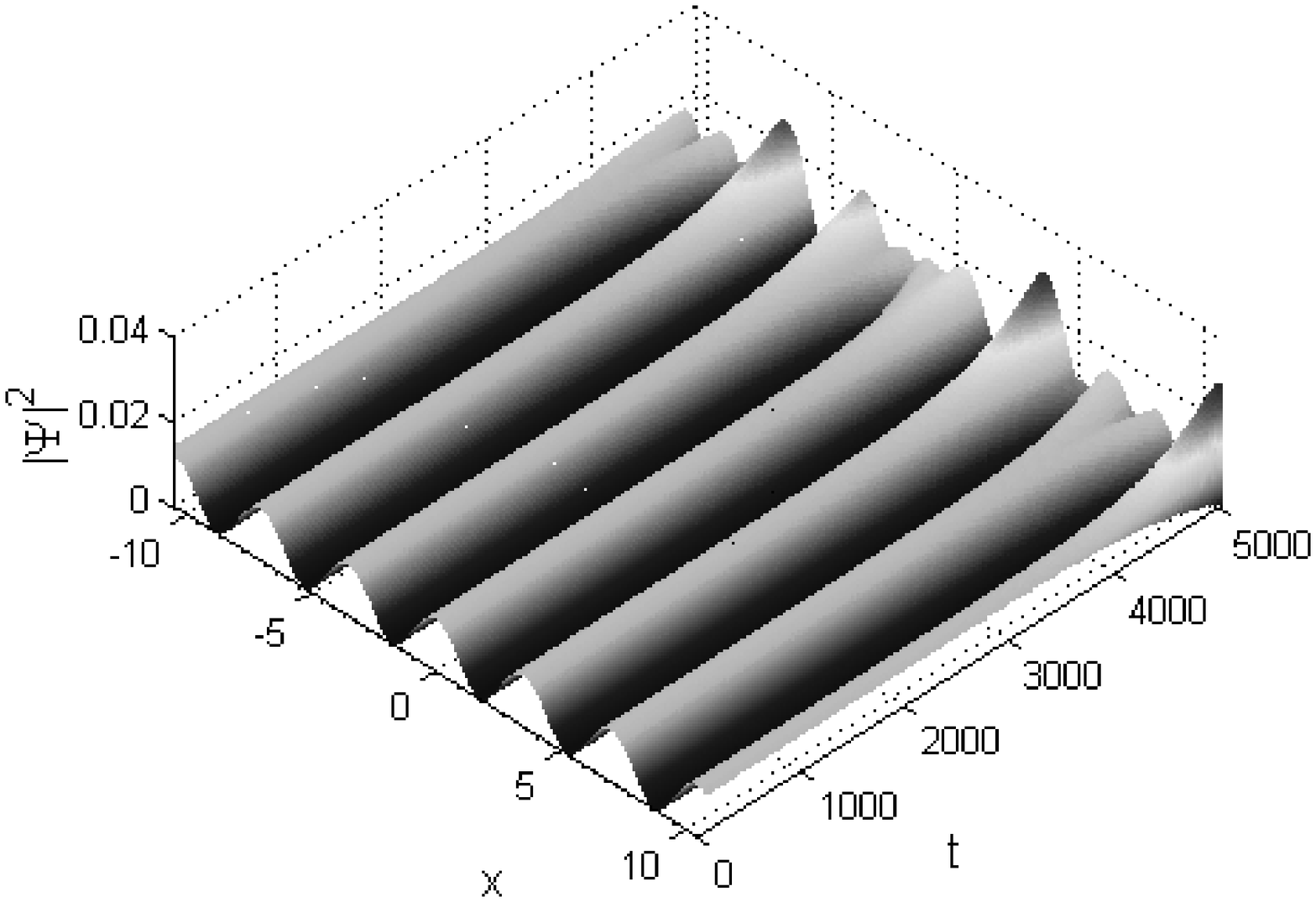}}
\par
\subfigure[]{\includegraphics [width=5.0cm]{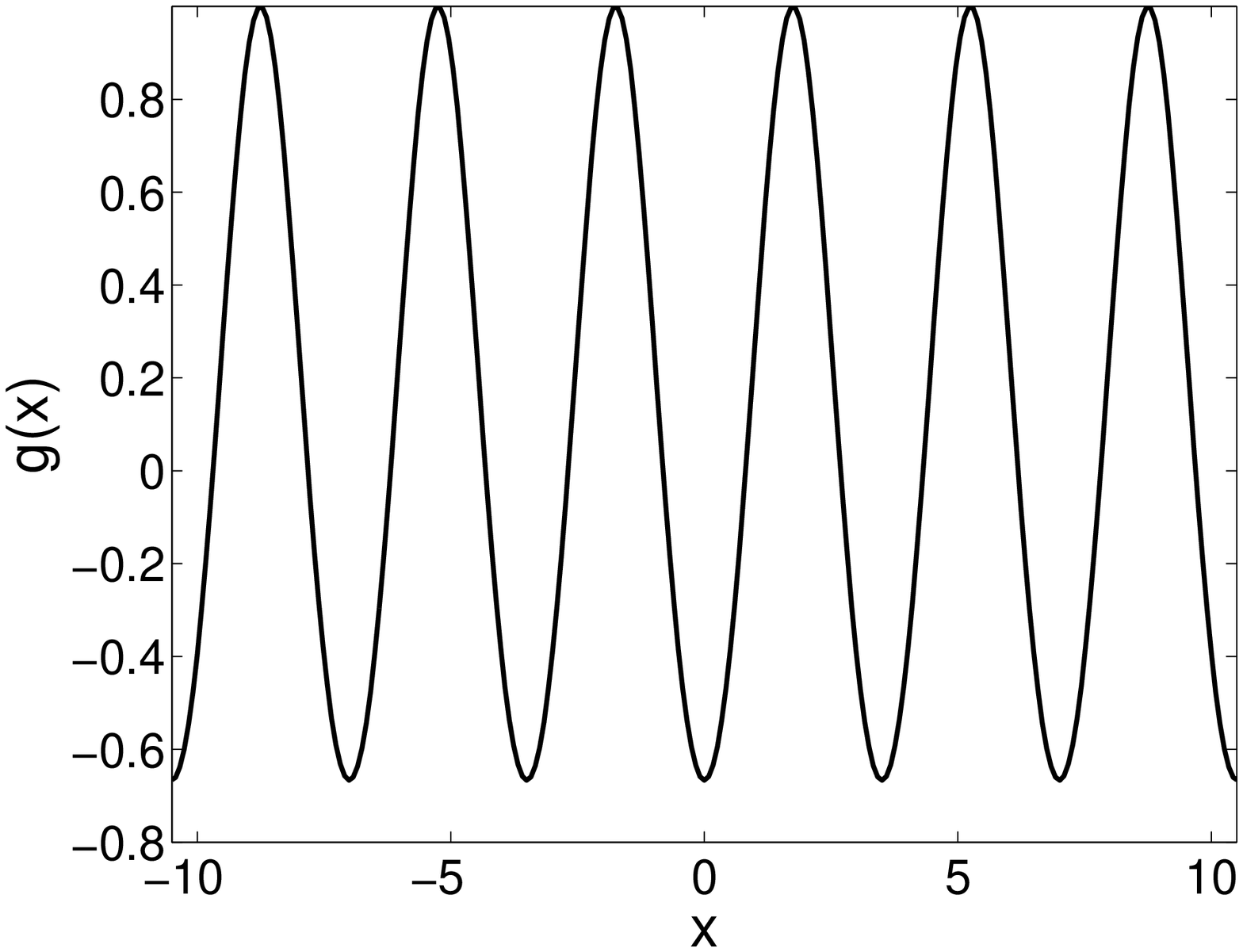}}%
\subfigure[]{\includegraphics [width=5.0cm]{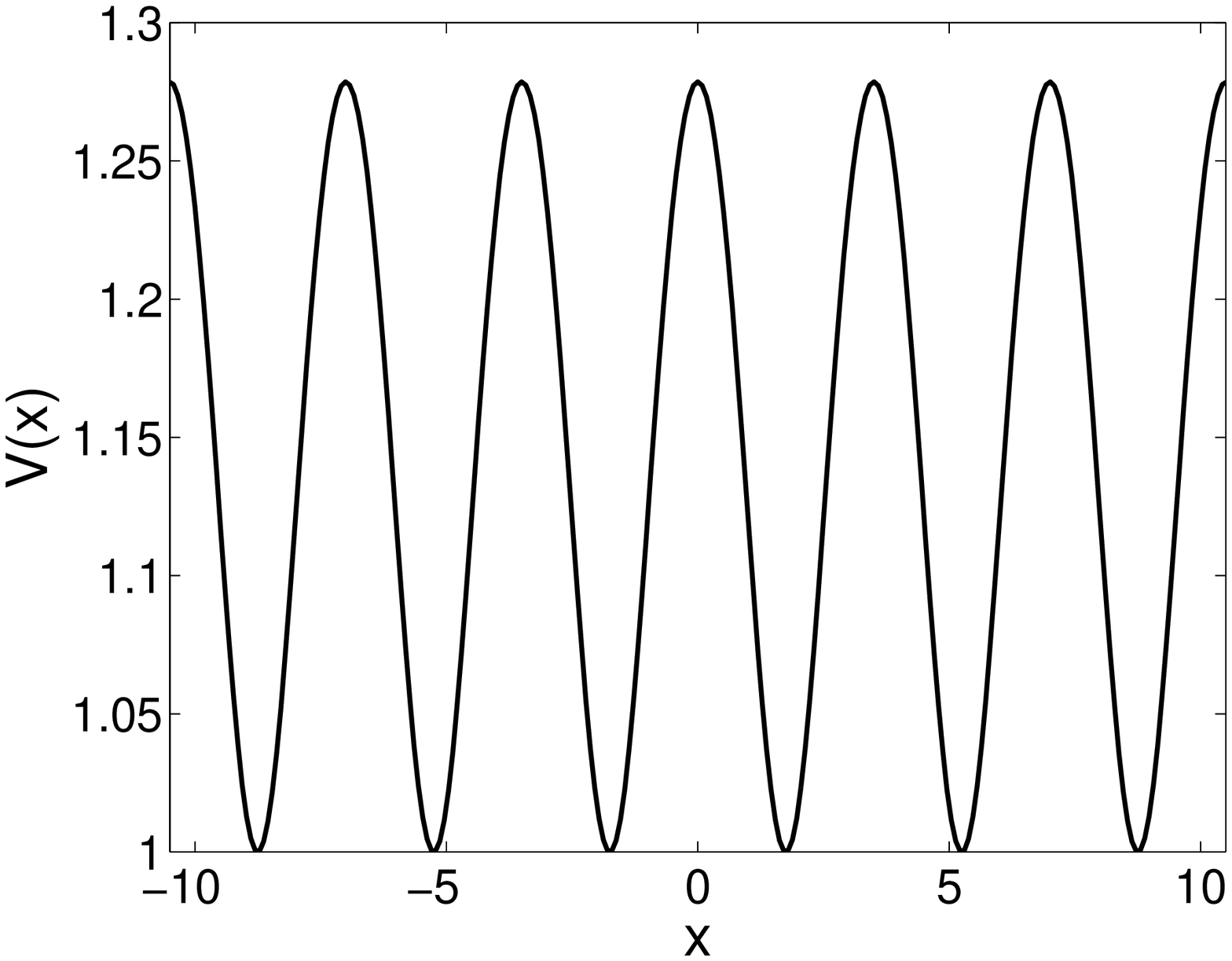}}
\caption{Panels (a) and (b) have the same meaning as in Figs. \protect\ref%
{fig1} - \protect\ref{fig3}, but here we display an example of the perturbed
evolution of a weakly unstable \textrm{cn}-type solution, found in the
presence of the linear periodic potential, as given by Eqs. (\protect\ref%
{V(x)}) and (\protect\ref{V0}). The shape of the linear potential is
displayed in panel (c). The parameters are $g_{0}=+1$, $b=+0.5$, $r=1$, $%
g_{1}=-2$, and $k=0.60$.}
\label{fig5}
\end{figure}

The linear potential given by expressions (\ref{V(x)}) and (\ref{V0})
affects not only the shape of the solutions but also their stability. First
of all, in the cases of $g_{0}=+1$ with either sign of $b$, and $g_{0}=-1$
with $b>0$, well-defined stable solutions have not been found, as well as in
the same cases in the absence of the linear potential, see above
(``well-defined" implies that we do not count solutions with a very small
amplitude, that may seem stable as the respective nonlinearity is extremely
weak). A typical example of a retarded onset of the instability of a
small-amplitude solution is displayed in Fig. \ref{fig5}, for $g_{0}=+1$ and
$b=+0.5$.

\begin{figure}[tbp]
\center\subfigure[]{\includegraphics [width=7.0cm]{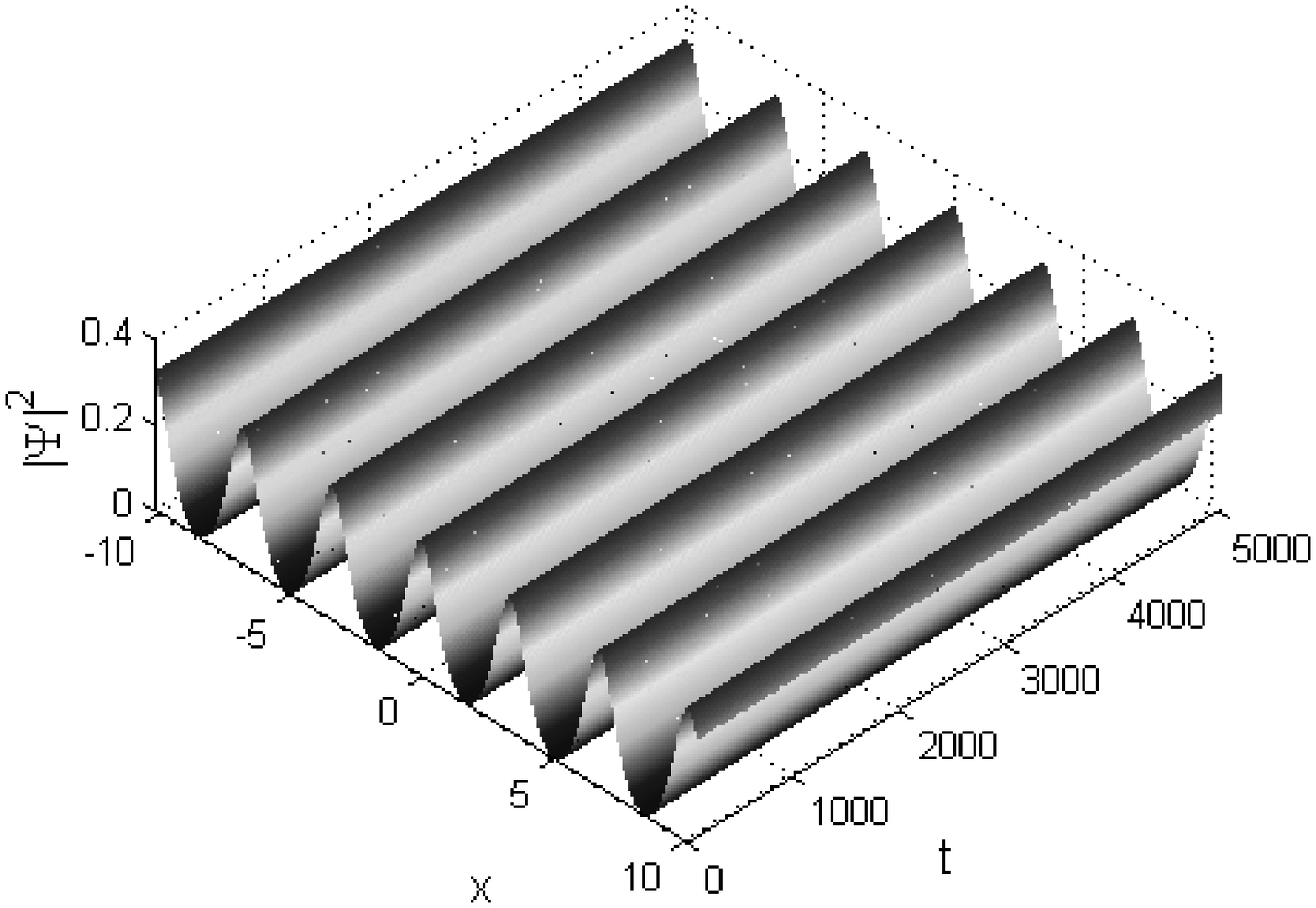}}
\par
\subfigure[]{\includegraphics [width=5.0cm]{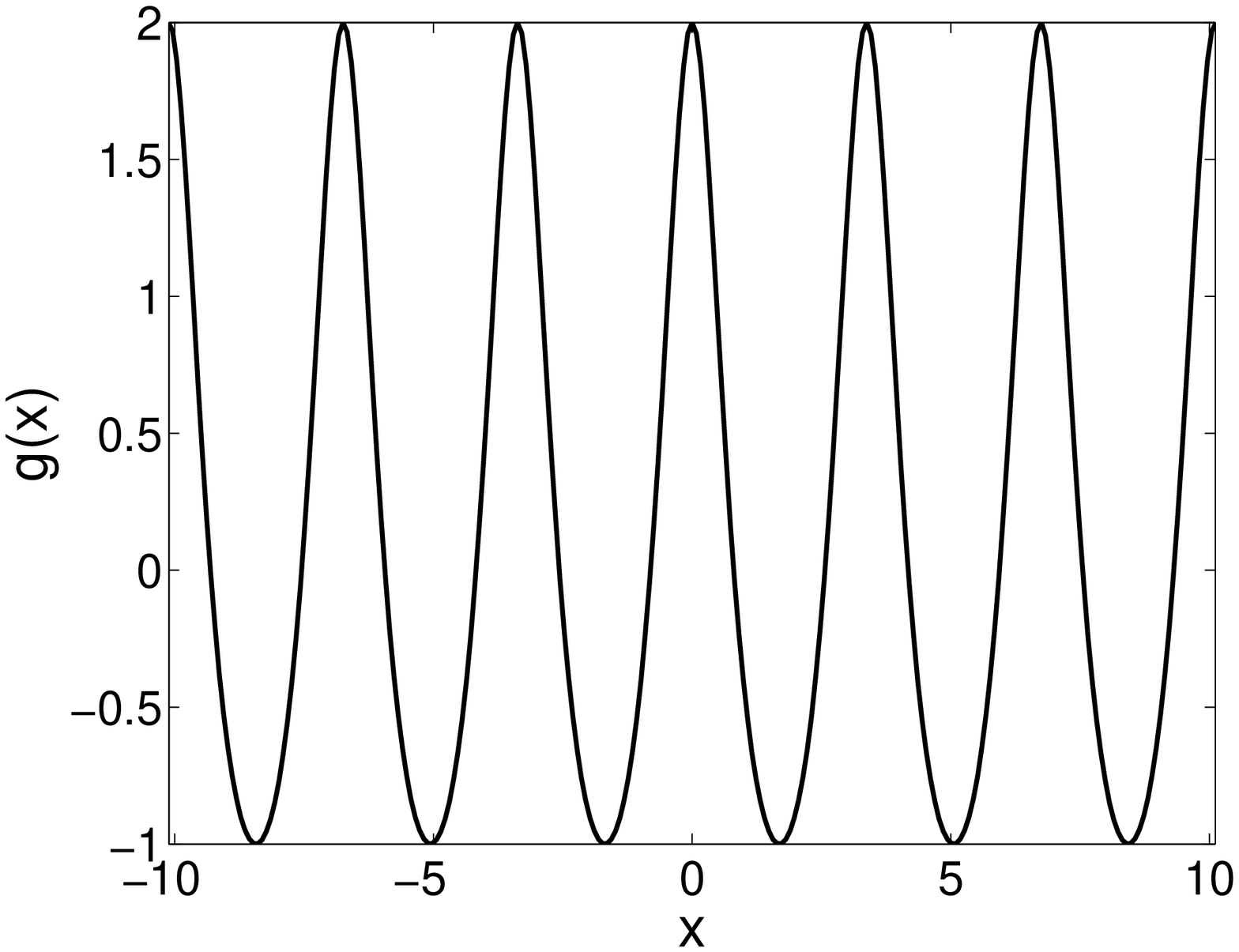}} %
\subfigure[]{\includegraphics [width=5.0cm]{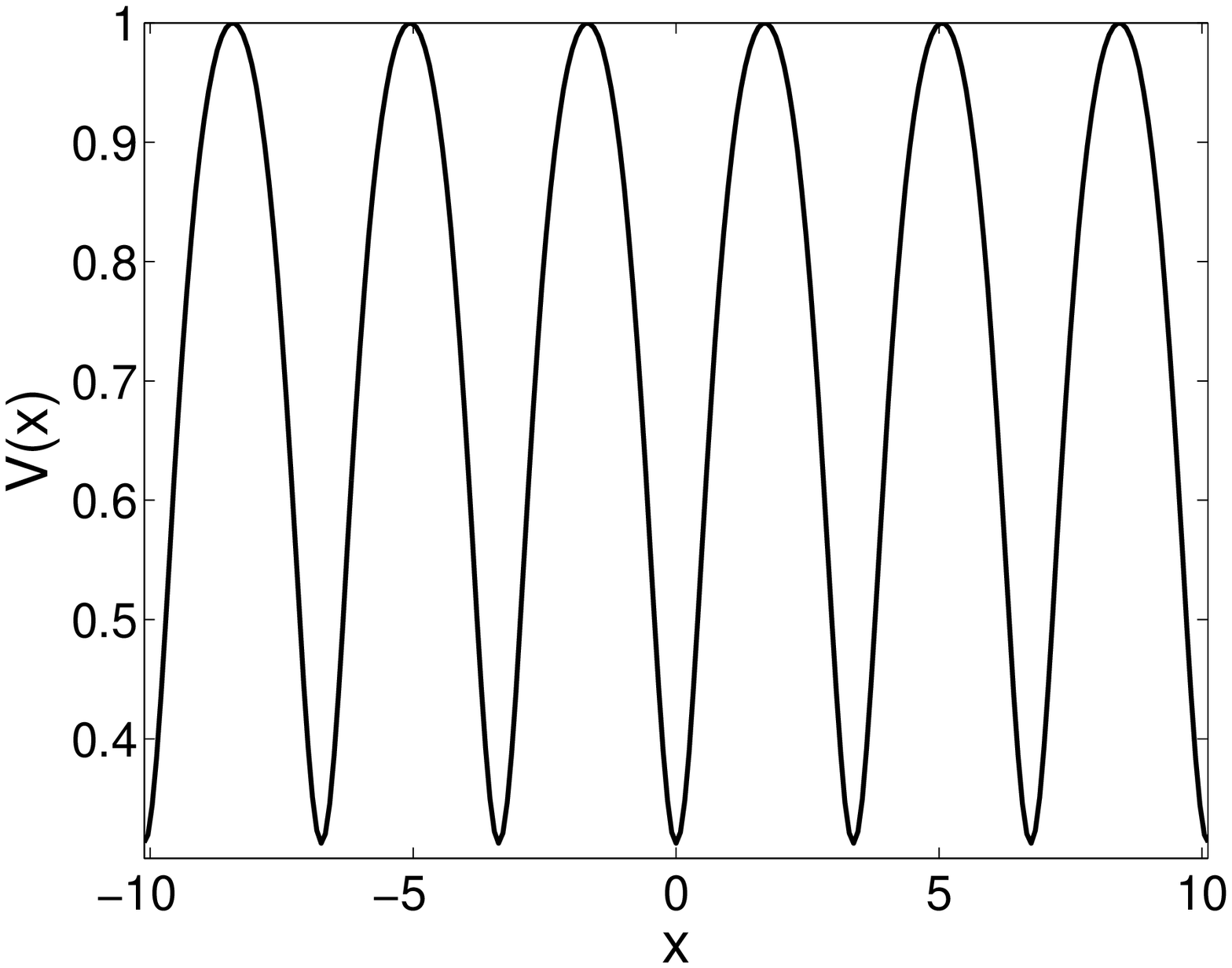}}
\caption{The same as in Fig. \protect\ref{fig5}, but for a stable
stationary solution found in the presence of the linear potential.
The parameters are $g_{0}=-1$, $b=-0.5$, $r=1$, $g_{1}=2$, and
$k=0.5$.} \label{fig6}
\end{figure}

\begin{figure}[tbp]
\center\subfigure[]{\includegraphics [width=7.0cm]{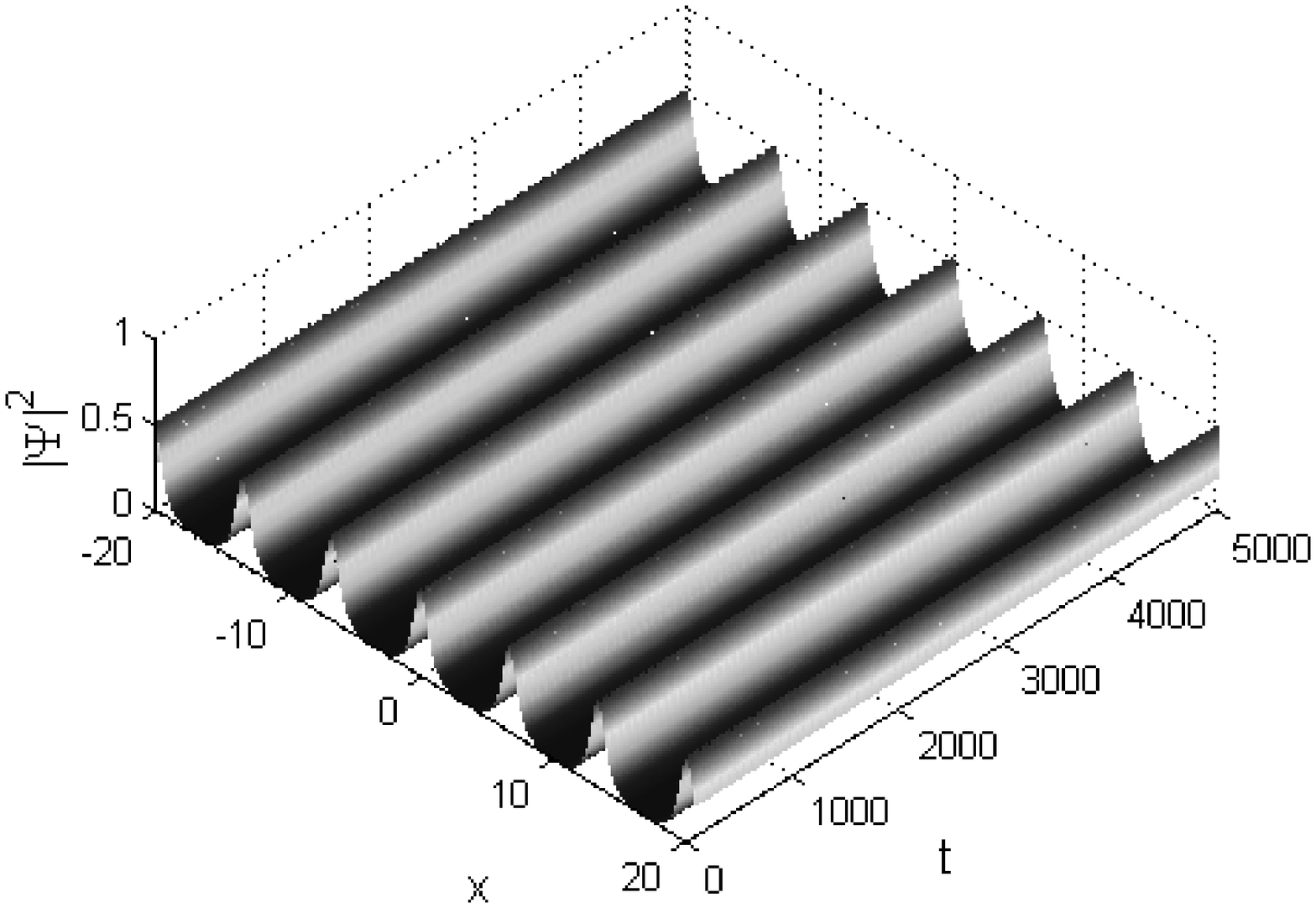}}
\par
\subfigure[]{\includegraphics [width=5.0cm]{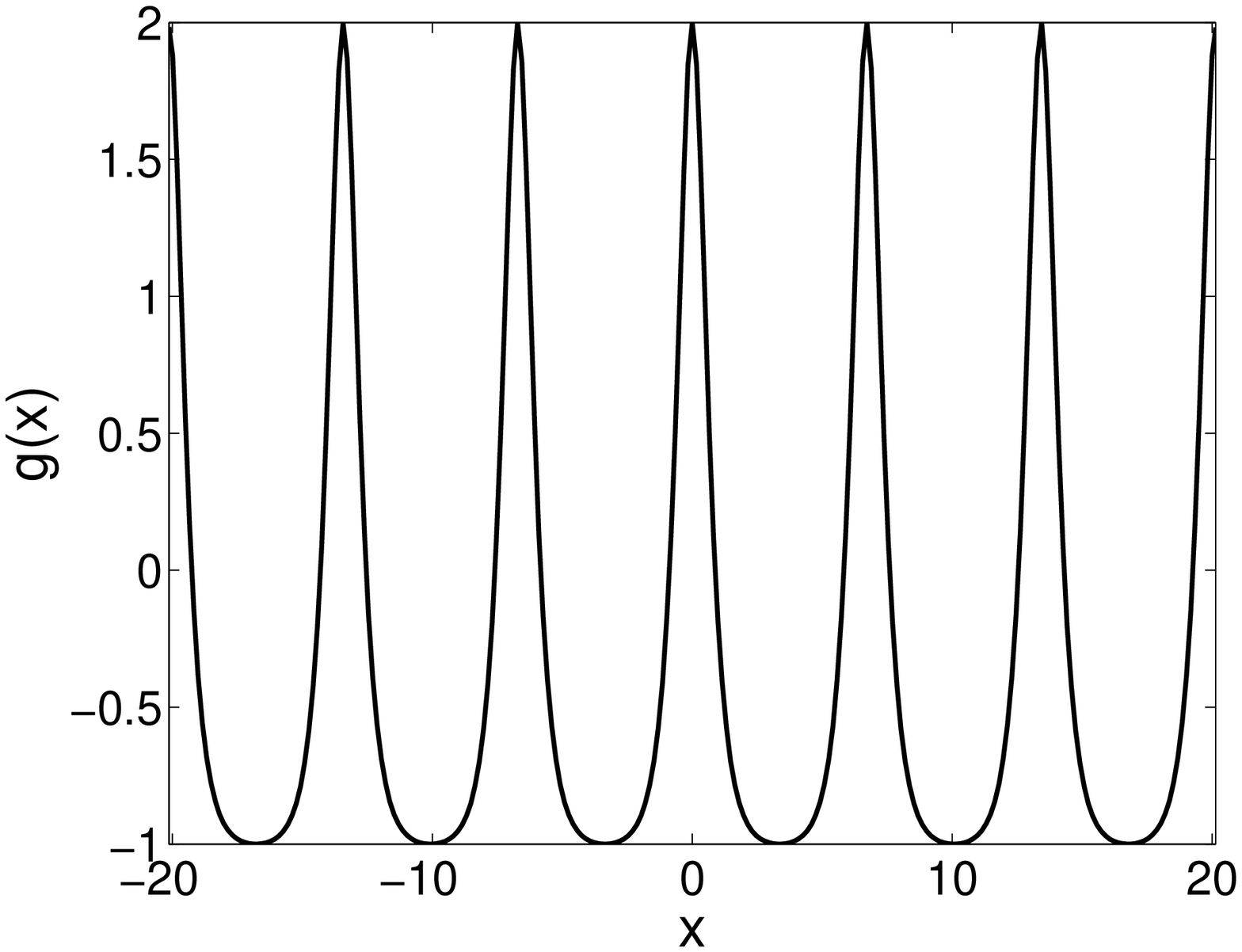}}%
\subfigure[]{\includegraphics [width=5.0cm]{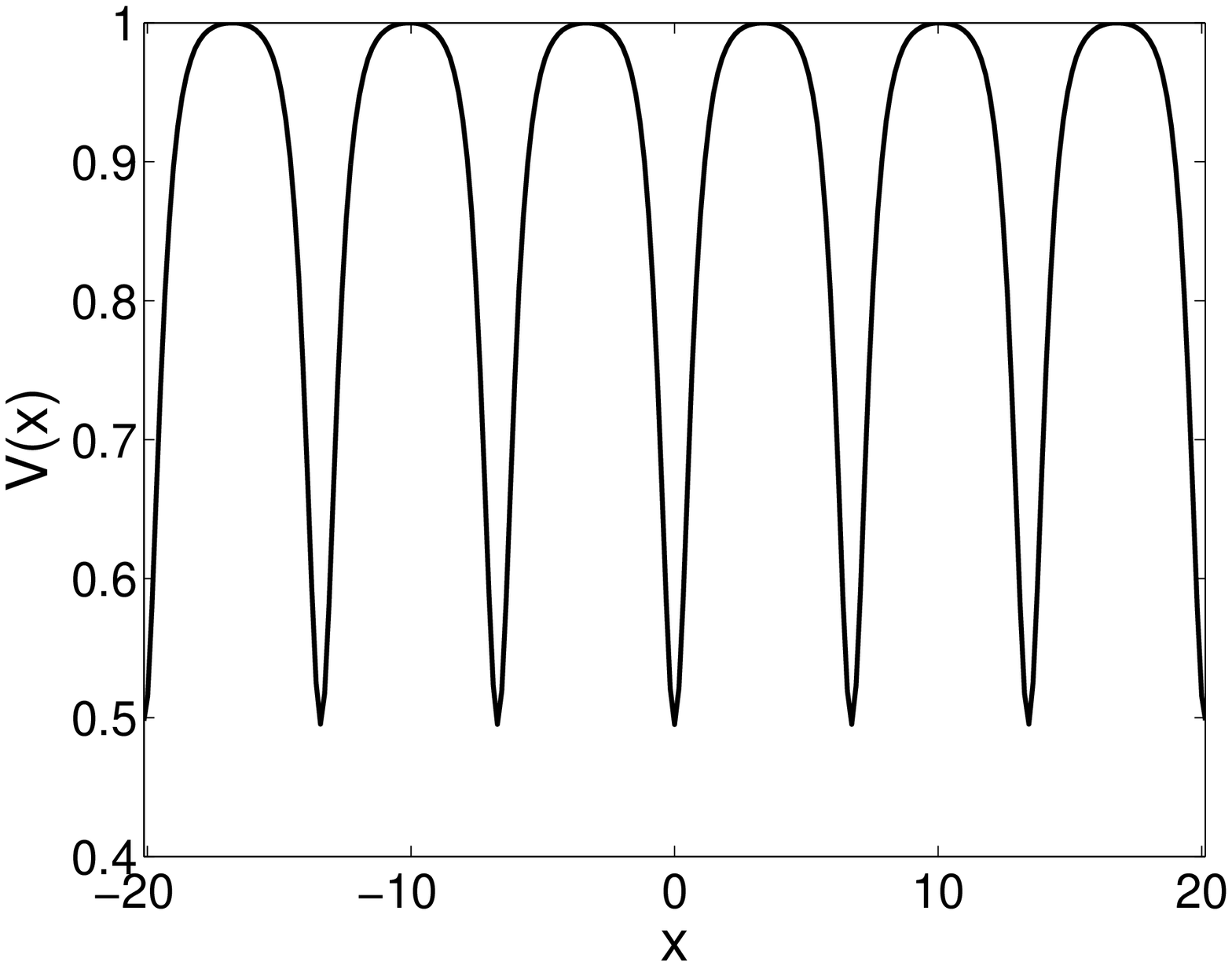}}
\caption{The same as in Fig. \protect\ref{fig6} (also a stable
solution), but for $k=0.99$.} \label{fig7}
\end{figure}

Similar to the situation reported above for the solutions with
$V(x)=0$, a vast subfamily of stable solutions is found in the case
of $g_{0}=-1$ and $b<0$. Actually, the stability region is located
at $b>-0.73$, this
border being practically independent of elliptical modulus $k$ (at $b=-0.72$%
, the exact solutions are found to be stable in interval $0.01\leq k\leq
0.99 $, while at $b=-0.74$ they are unstable in the same interval). The
shape of the stable solutions is quasi-harmonic at small and moderate values
of $k$ (see an example in Fig. \ref{fig6}), while featuring sharp peaks at $%
k $ close to $1$ (Fig. \ref{fig7}). Note that the amplitude of the solutions
shown in these typical examples is not especially small, which confirms that
the observed stability is not a trivial consequence of the weak
nonlinearity. It is also worthy to note that density maxima of the solutions
coincide with maxima of the NL modulation function, $g(x)$, and minima of $%
V(x)$, i.e., the density maxima are collocated with minima of \emph{both }%
the nonlinear pseudopotential and linear potential. This circumstance
obviously facilitates the stabilization of the periodic patterns, giving
them a chance to realize the ground state.

\section{Exact solutions of \textrm{dn} type and their stability}

\subsection{Stationary solutions}

Another generic class of exact solutions, which, as well as the solutions of
the \textrm{cn} type, are represented by even functions of $x$, can be
obtained from \textrm{cn} by the continuation to values of $k>1$, using the
well-known transformation formula,
\begin{equation}
\mathrm{cn}\left( x,k\right) =\mathrm{dn}\left( kx,1/k\right) .  \label{k}
\end{equation}%
Thus, in the absence of the linear potential, $V(x)=0$, expressions (\ref{g}%
)-(\ref{mu}) generate the following class of exact solutions to Eq. (\ref%
{psi}):%
\begin{equation}
g(x)=\frac{g_{0}+g_{1}\mathrm{dn}^{2}(rx)}{1+b\hspace{0.05in}\mathrm{dn}%
^{2}(rx)},  \label{gdn}
\end{equation}%
\begin{equation}
\psi (x)=A_{0}\frac{\mathrm{dn}(rx)}{\sqrt{1+b\hspace{0.05in}\mathrm{dn}%
^{2}(rx)}},  \label{solutiondn}
\end{equation}%
\begin{equation}
g_{1}=\frac{g_{0}b}{2}\frac{\left( b+1\right) (3b-1)-b(3b+1)k^{2}}{\left(
b+1\right) (3b+1)-b(3b+2)k^{2}},  \label{g1dn}
\end{equation}%
\begin{equation}
A_{0}^{2}=g_{0}r^{2}[\left( b+1\right) (3b+1)-b(3b+2)k^{2}],  \label{Adn}
\end{equation}%
\begin{equation}
\mu =\left( r^{2}/2\right) [-2-3b+(3b+1)k^{2}],  \label{mudn}
\end{equation}%
where, as above, we assume normalization $g_{0}=\pm 1$, the elliptic modulus
takes values $0<k<1$, and an additional normalization may be imposed, $r=1$.

As concerns the transition to the ordinary GPE, with $g=\mathrm{const}$, Eq.
(\ref{gdn}) demonstrates that this is possible in two cases, $b=0$ or under
condition coinciding with Eq. (\ref{1b0}), see above. The latter condition,
if applied to Eq. (\ref{g1dn}), leads to $k^{2}=\left( b+1\right) /b$.
Unlike the similar condition obtained for the \textrm{cn}-type solutions in
the form \ of Eq. (\ref{b/(1+b)}), the present one, obviously, cannot give
appropriate values of the elliptical modulus, $0<k^{2}<1$, if $b$ takes
values for which solution (\ref{solutiondn}) is nonsingular, i.e., $b>-1$.
As for the former limit case, $b=0$, expressions (\ref{solutiondn}) and (\ref%
{Adn}) yield the ordinary unstable solution to the equation with the uniform
self-attraction, $g(x)\equiv +1$, in the form of $\psi (x)=\mathrm{dn}(x)$.

Making use of Eqs. (\ref{V(x)})-(\ref{muV}) taken at $k>1$ and reducing this
to values $k<1$ by means of Eq. (\ref{k}), the \textrm{dn} solution family
may be extended to the case when the linear potential is present in Eq. (\ref%
{psi}):
\begin{equation}
V(x)=\frac{V_{0}\mathrm{dn}^{2}(rx)}{1+b~\mathrm{dn}^{2}(rx)},
\label{V(x)dn}
\end{equation}%
\begin{gather}
V_{0}=\frac{r^{2}}{2}+\frac{br^{2}}{2}[(3b+1)k^{2}-\left( 3b+2\right) ]
\notag \\
-\frac{3(b+1)g_{1}}{2(g_{0}b-g_{1})}[bk^{2}-(1+b)],  \label{V0dn}
\end{gather}%
\begin{equation}
A_{0}^{2}=\frac{3br^{2}(1+b)[bk^{2}-(1+b)]}{2\left( g_{1}-g_{0}b\right) }~,
\label{AVdn}
\end{equation}

\begin{equation}
\mu =(r^{2}/2)[(3b+1)k^{2}-(2+3b)].  \label{muVdn}
\end{equation}%
Similar to the class of the \textrm{cn}-based solutions, in the presence of
the linear potential the NL-modulation function is given by the same general
expression as above, Eq. (\ref{gdn}), while Eq. (\ref{g1dn}) should be
dropped (actually, the latter equation is the condition for reducing $V_{0}$
to zero). Accordingly, the solution family depends on four free parameters, $%
g_{1},r,k,b$, and the sign parameter, $g_{0}$.

\subsection{The stability of the \textrm{dn}-type solutions}

\begin{figure}[tbp]
\center\subfigure[]{\includegraphics [width=7.0cm]{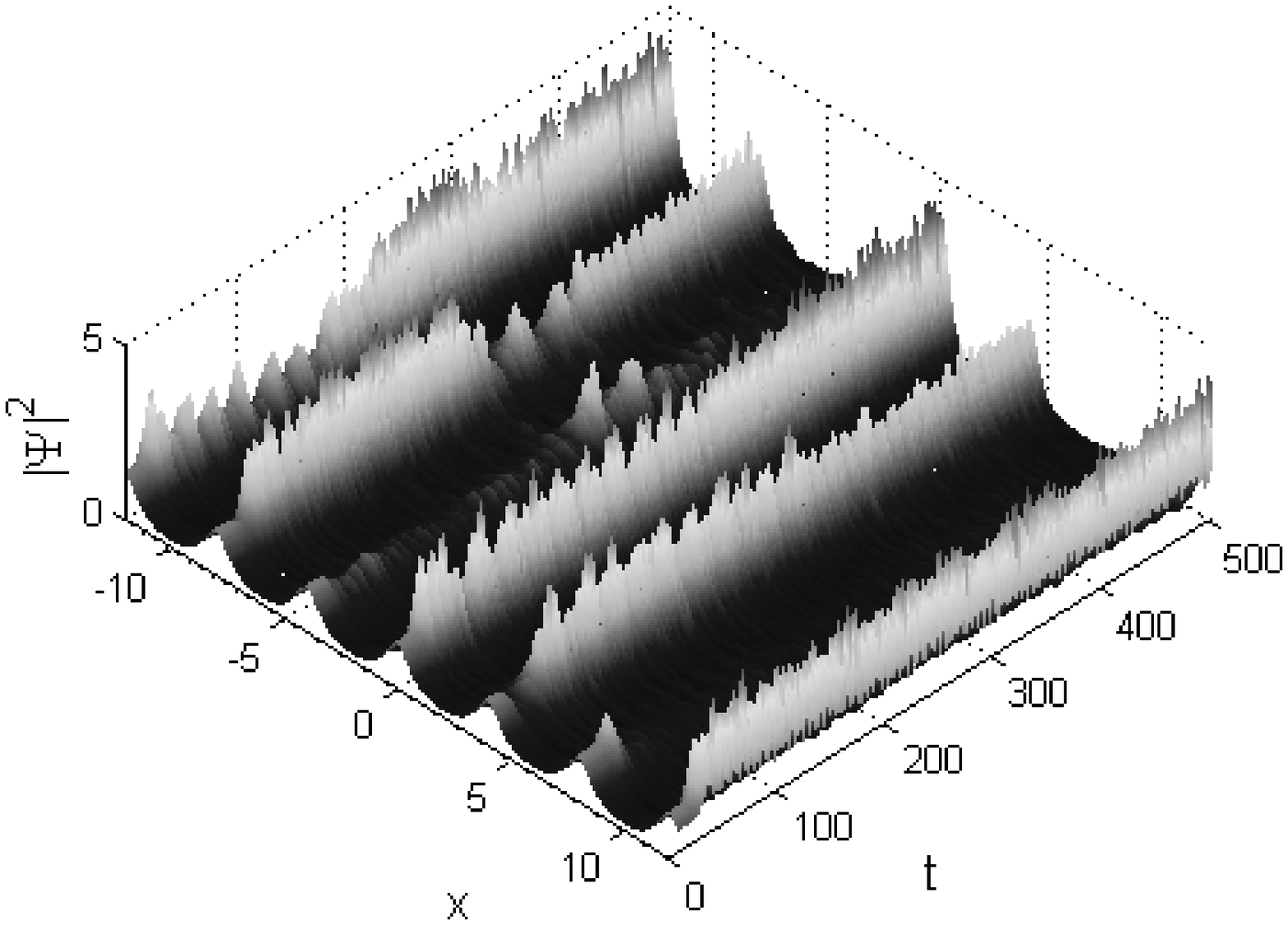}}%
\subfigure[]{\includegraphics [width=5.0cm]{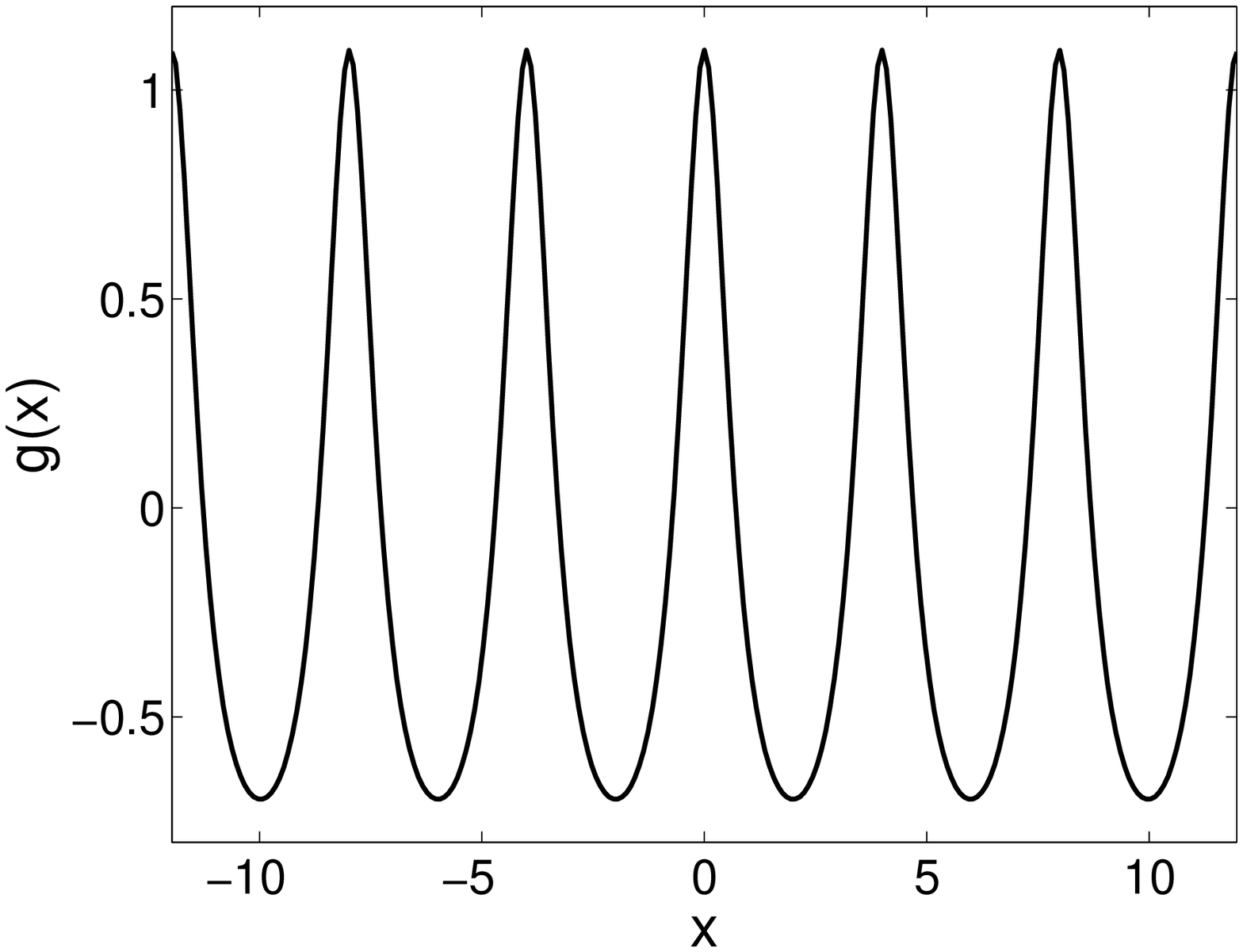}}
\caption{The same as in Fig. \protect\ref{fig5}, but for a typical
unstable solution of the \textrm{dn} type found in the absence of
the linear potential, with $g_{0}=-1$, $b=-0.7$, $r=1$,
$g_{1}=1.33$, and $k=0.8$.} \label{fig8}
\end{figure}

In the absence of the linear potential, the exact solutions given by Eqs. (%
\ref{gdn})-(\ref{Adn}) are found to be strongly unstable, see a typical
example in Fig. \ref{fig8}. Also unstable are the exact solutions found in
the presence of the linear potential, as per Eqs. (\ref{V(x)dn})-(\ref{AVdn}%
), in the case of $g_{0}=-1$, on the contrary to the solutions of the
\textrm{cn} type, and also to the odd solutions of the \textrm{sn} type (see
below), where the potential can readily stabilize the solutions with $%
g_{0}=-1$.

\begin{figure}[tbp]
\center\subfigure[]{\includegraphics [width=7.0cm]{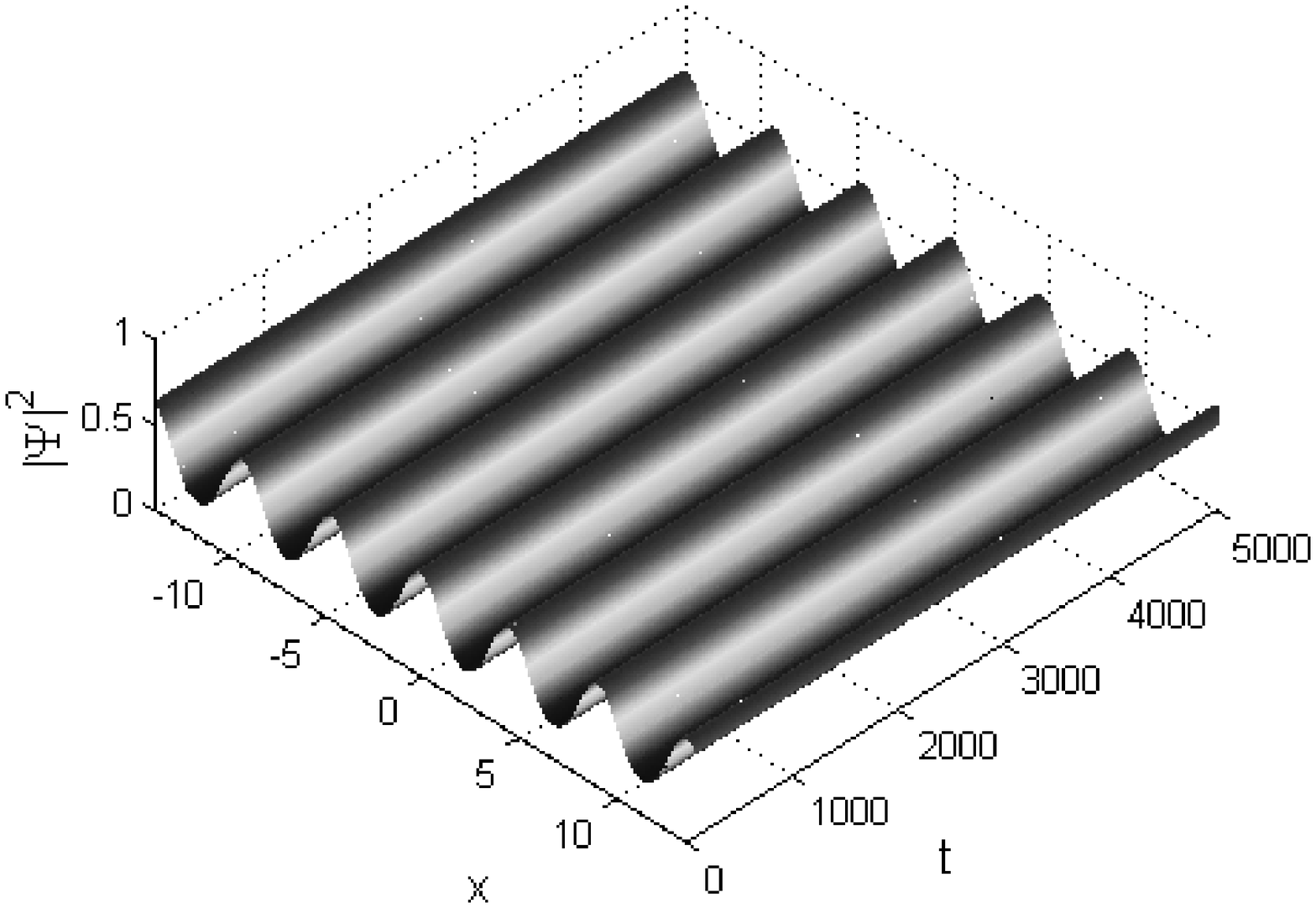}}
\par
\subfigure[]{\includegraphics [width=5.0cm]{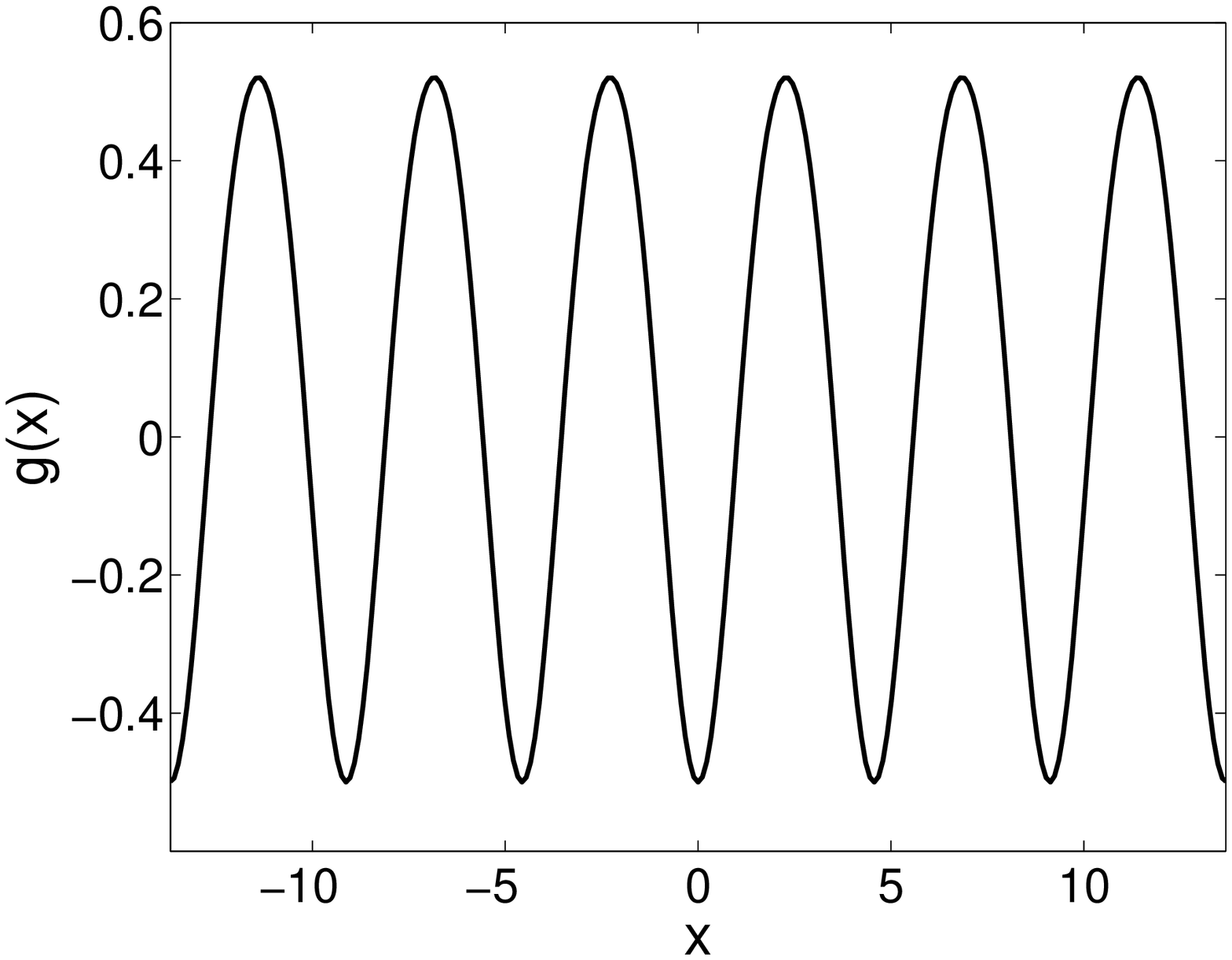}}%
\subfigure[]{\includegraphics [width=5.0cm]{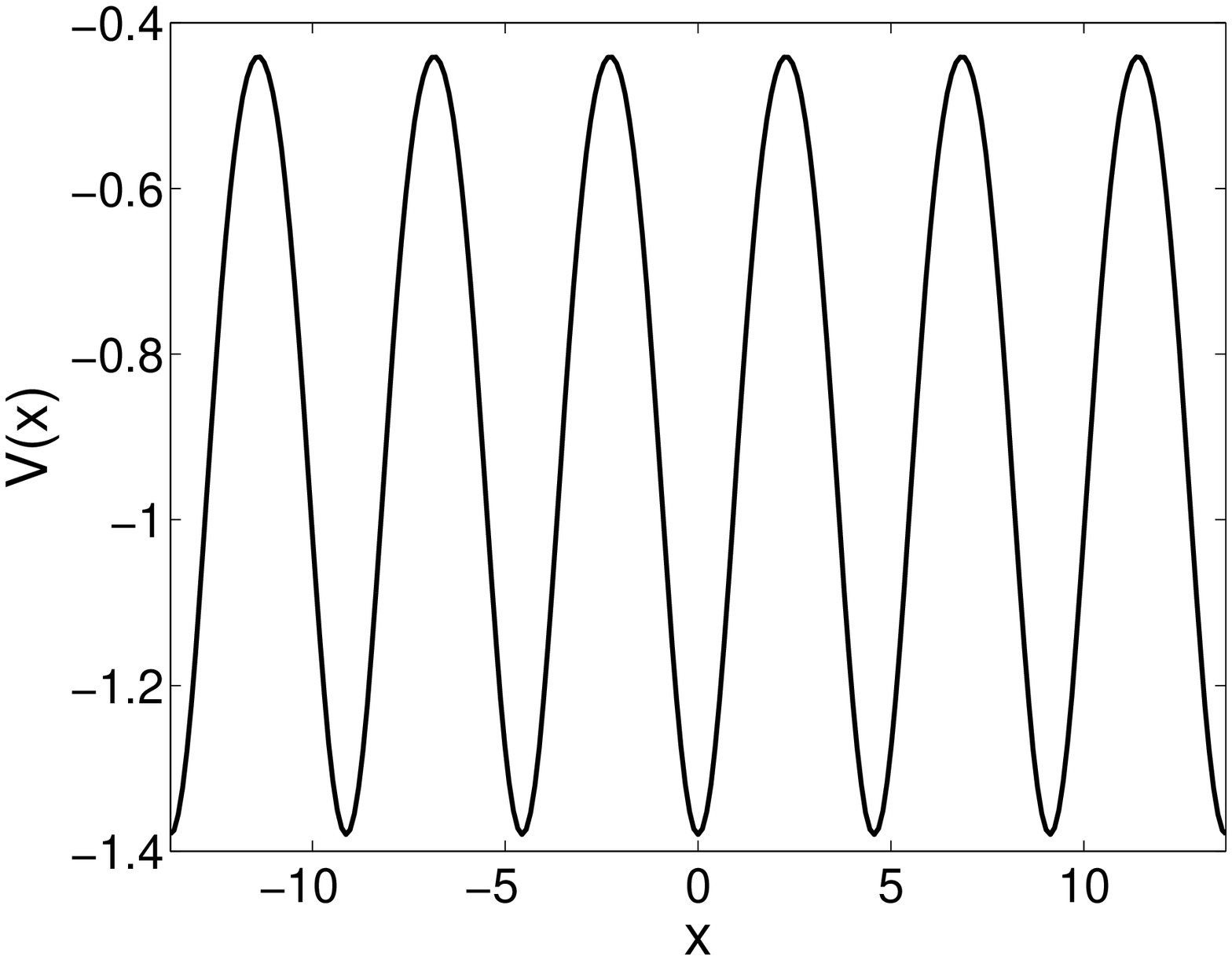}}
\caption{The same as in Fig. \protect\ref{fig8}, but for a typical \emph{%
stable} solution of the \textrm{dn} type found in the presence of the linear
potential, for $g_{0}=+1$, $b=1$, $r=1$, $g_{1}=-2$, and $k=0.9$.}
\label{fig9}
\end{figure}

On the other hand, in the case of $g_{0}=+1,$ when the solutions of both the
\textrm{cn} (see above) and \textrm{sn} (see below) types cannot be stable,
the \textrm{dn} species of the periodic patterns readily demonstrates its
stability, in the presence of the linear potential, as shown in Fig. %
\ref{fig9}. Note that panels (b) and (c) of this figure imply the \textit{%
competition} between the nonlinear pseudopotential and linear potential, as
minima of $g(x)$ coincide with minima of $V(x)$, this sets of points also
coinciding with density maxima of the solution, see panel (a). A similar
competition between the nonlinear and linear effective potentials was
recently considered in a model of a one-dimensional photonic crystal \cite%
{Thawatachai}. In fact, almost all the solutions belonging to this subfamily
are stable, with the exception of ones with large $b$ or $k$ very close to $%
1 $.

\section{Solutions of the \textrm{sn} type and their stability}

\subsection{Solutions without the linear potential}

In addition to the two classes of even stationary solutions considered
above, it is possible to find a class of the NL-modulation functions, $g(x)$%
, and related potentials, $V(x)$, which are even too, but they
support exact stationary solutions which are \emph{odd} in $x$, that
are based on the Jacobi's function $\mathrm{sn}$. First, in the case
of $V(x)=0$, the explicit form of these solutions, along with
function $g(x)$ which supports
them, is [cf. similar expressions (\ref{g})-(\ref{mu}) and (\ref{gdn})-(\ref%
{mudn}) for the solutions of the \textrm{cn} and \textrm{dn} types]:
\begin{equation}
g(x)=\frac{g_{0}+g_{1}\mathrm{sn}^{2}(rx)}{1+b\hspace{0.05in}\mathrm{sn}%
^{2}(rx)},  \label{gsn}
\end{equation}%
\begin{equation}
\psi =\frac{A_{0}\mathrm{sn}(rx)}{\sqrt{1+b\hspace{0.05in}\mathrm{sn}^{2}(rx)%
}},  \label{solutionsn}
\end{equation}%
\begin{equation}
g_{1}=\frac{g_{0}b}{2}\frac{b(1+3b)-(1-b)k^{2}}{b(2+3b)+(1+2b)k^{2}},
\label{g1sn}
\end{equation}%
\begin{equation}
A_{0}^{2}=-g_{0}r^{2}[b(2+3b)+(1+2b)k^{2}],  \label{A0sn}
\end{equation}%
\begin{equation}
\mu =\frac{r^{2}}{2}\left( 1+3b+k^{2}\right) .  \label{musn}
\end{equation}

In the case of $g_{0}=+1$, expression (\ref{A0sn}) cannot give $A_{0}^{2}>0$
for $b>0$. Further, a straightforward analysis demonstrates that, in the
same case but for $b$ taken from $0<-b<1/3$, the solutions exist in interval%
\begin{equation}
k^{2}<\frac{|b|\left( 2-3|b|\right) }{1-2|b|}  \label{k^21}
\end{equation}%
of values of the elliptic modulus. This entire interval satisfies
constraint $k<1$. For $1/3<-b<2/3$, the solutions exist for all
$k^{2}<1$. Finally,
for $2/3<-b<1$, the existence condition is%
\begin{equation}
\frac{|b|\left( 3|b|-2\right) }{2|b|-1}~<k^{2}<1.  \label{k^22}
\end{equation}

A similar analysis performed for the case of $g_{0}=-1$ demonstrates that,
with $b>0$, the solutions exist for all values of $k^{2}<1$. Proceeding to
negative $b$ in this case, it is easy to conclude that, for $1/3<-b<0$, the
solutions with $A_{0}^{2}>0$ exist for%
\begin{equation}
\frac{|b|\left( 2-3|b|\right) }{1-2|b|}<k^{2}<1,
\end{equation}%
cf. existence condition (\ref{k^21}) for $g_{0}=+1$, which was obtained in
the same interval of $b$. In the adjacent interval, $1/3<-b<2/3$, the
solutions do not exist at all. Finally, in the end portion of the allowed
interval of $b$, \textit{viz}., $2/3<-b<1$, the \textrm{sn}-type solutions
are available in the region of%
\begin{equation}
k^{2}<\frac{|b|\left( 3|b|-2\right) }{2|b|-1},  \label{last}
\end{equation}%
cf. region (\ref{k^22}) in the case of $g_{0}=+1$ [the whole interval (\ref%
{last}) complies with constraint $k<1$].

The transition of $g(x)$ given by Eq. (\ref{gsn}) to
$g=\mathrm{const}$, corresponding to the ordinary GPE, may be
performed by taking $b=0$, or by imposing the same condition
(\ref{1b0}) as above. In the former case, the respective solution,
$\psi (x)=k~\mathrm{sn}(x)$, satisfies the GPE with $g(x)\equiv -1$.
In the
latter case, the solution is relevant for $g(x)\equiv +1$, reducing to $%
b=-k^{2}$, $\mu =(1/2)\left( 1-2k^{2}\right) $, and $\psi (x)=k\sqrt{1-k^{2}}%
\mathrm{sn}(x)/\mathrm{dn}(x)$, cf. similar limit solution (\ref{limit}) of
the \textrm{cn} type.

\subsection{Exact \textrm{sn} solutions with a linear potential}

Another family of the \textrm{sn} solutions can be found, adding to Eq. (\ref%
{psi}) the linear potential in the following form,%
\begin{equation}
V(x)=-\frac{U_{0}\mathrm{sn}^{2}(rx)}{1+b~\mathrm{sn}^{2}(rx)},  \label{Vsn}
\end{equation}%
\begin{equation}
U_{0}=\frac{k^{2}r^{2}}{2}-\frac{b(1+3b+k^{2})r^{2}}{2}-\frac{%
3g_{1}(1+b)(b+k^{2})r^{2}}{2(g_{1}-g_{0}b)}.  \label{U0}
\end{equation}%
Then, $\psi (x)$ is again given Eq. (\ref{solutionsn}), but Eqs. (\ref{A0sn}%
) and (\ref{musn}) are replaced by
\begin{equation}
A_{0}^{2}=\frac{3b(1+b)(b+k^{2})r^{2}}{2(g_{1}-g_{0}b)},  \label{AVsn}
\end{equation}%
\begin{equation}
\mu =\left( r^{2}/2\right) (1+3b+k^{2}),  \label{muVsn}
\end{equation}%
while Eq. (\ref{g1sn}) is dropped.

In the limit of $b\rightarrow 0$ and $g_1 \rightarrow 0$, the present solution makes sense for $%
g_{0}=-1$. In this limit, we may again fix $r=1$, hence Eqs. (\ref%
{solutionsn}), (\ref{AVsn}) and (\ref{muVsn}) yield
\begin{equation}
\psi =\sqrt{3/2}k~\mathrm{sn}(x),~\mu =\frac{1+k^{2}}{2}.  \label{old}
\end{equation}%
Expressions (\ref{old}) provide for a solution to an equation with the
constant nonlinearity coefficient,
\begin{equation}
\mu \psi +(1/2)\psi ^{\prime \prime }-\psi ^{3}+\left( k^{2}/2\right)
\mathrm{sn}^{2}(x)\psi =0\text{.}
\end{equation}%
These particular results, obtained for $g(x)\equiv -1$, reproduce findings
reported in Ref. \cite{Carr1}.

\subsection{The stability}

\begin{figure}[tbp]
\center\subfigure[]{\includegraphics [width=7.0cm]{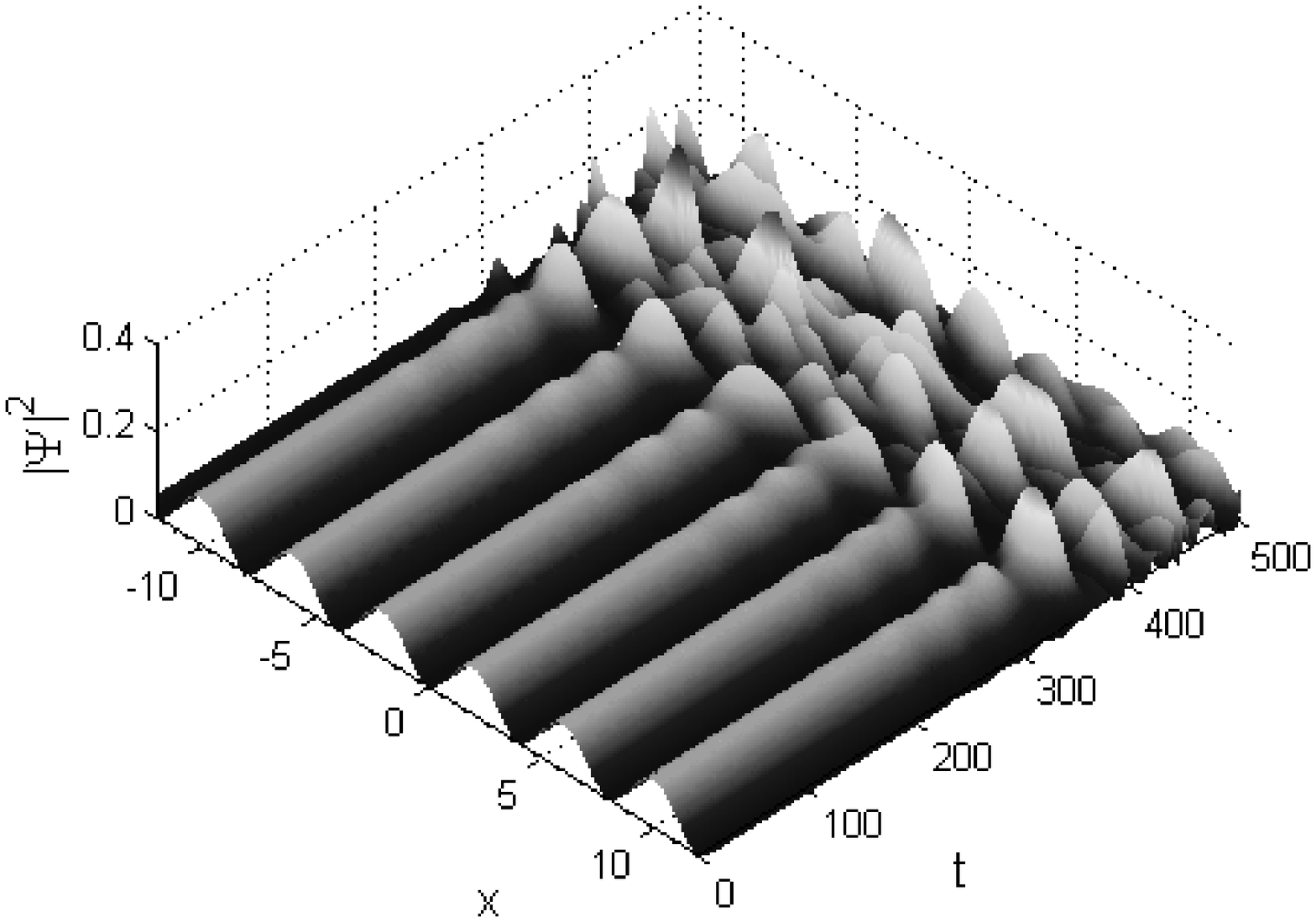}}%
\subfigure[]{\includegraphics [width=5.0cm]{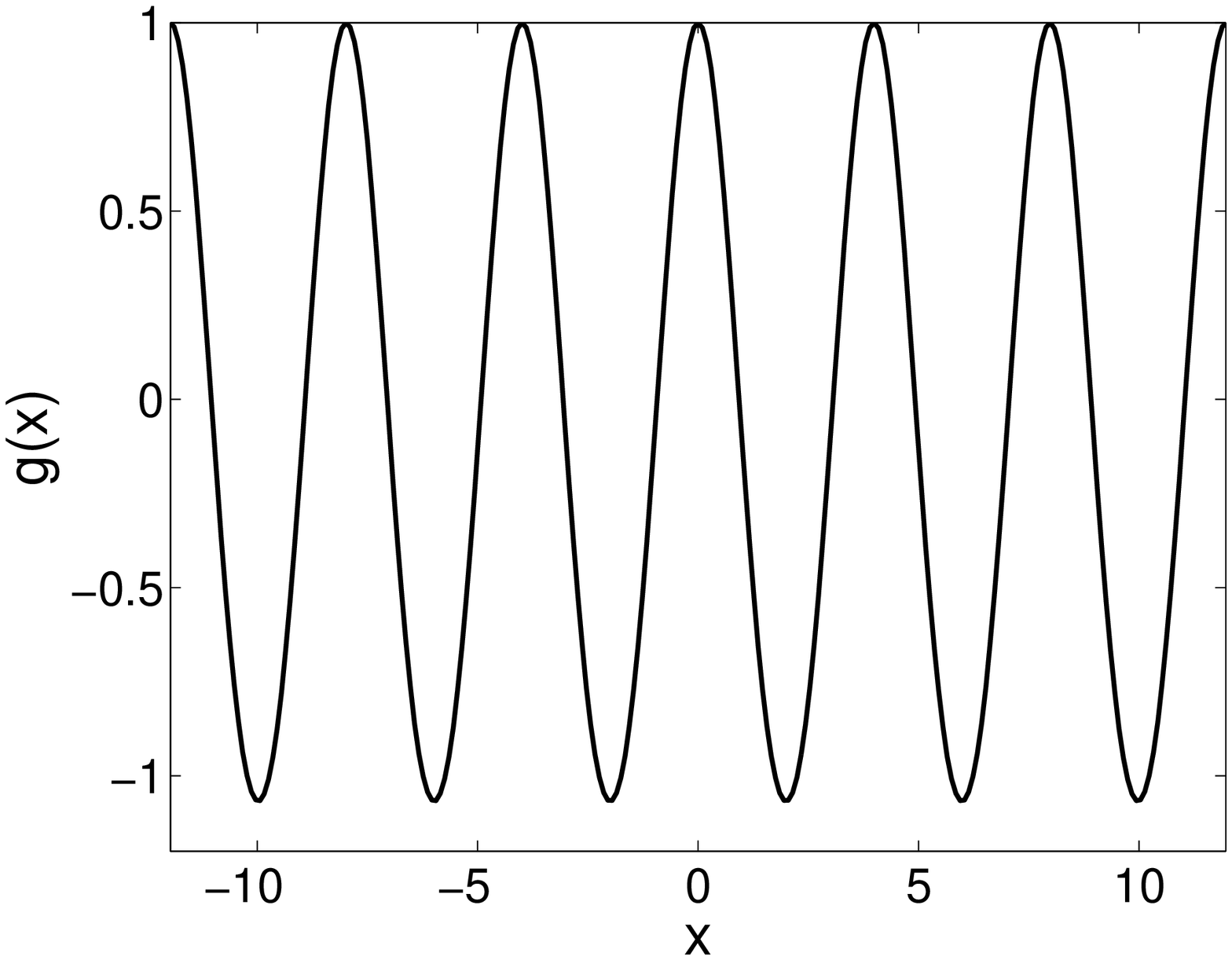}}
\caption{The same as in Figs. \protect\ref{fig1},
\protect\ref{fig3}, and \protect\ref{fig8}, but for a typical
unstable solution of the \textrm{sn}
type in the absence of the linear potential. The parameters are $g_{0}=+1$, $%
b=-0.3$, $r=1$, $g_{1}=-1.75$, and $k=0.8$.}
\label{fig10}
\end{figure}

In the absence of the linear potential, nearly all the solutions of the
\textrm{sn} type are unstable, as illustrated by a typical example in Fig. %
\ref{fig10}. Note that, in this case, maxima of the density coincide with
minima of $g(x)$, thus pushing the solutions towards instability. The only
stable solutions could be found for $g_{0}=-1$ and very small $|b|$, when,
as said above, the exact solutions are close to the commonly known stable
exact solution, $\psi (x)=k~\mathrm{sn}(x)$, to the GPE with $g(x)\equiv -1$%
.

\begin{figure}[tbp]
\center\subfigure[]{\includegraphics [width=7.0cm]{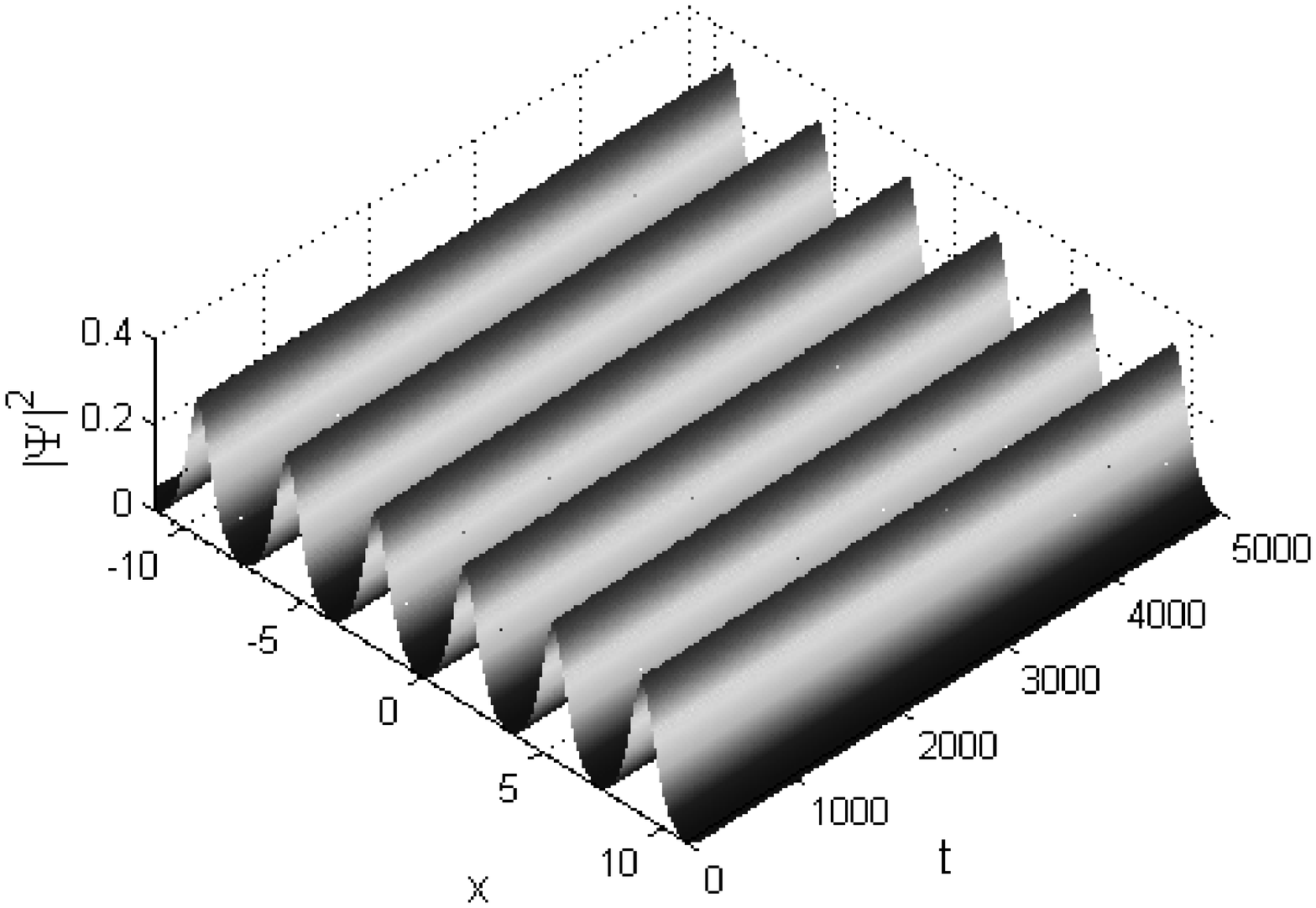}}
\par
\subfigure[]{\includegraphics [width=5.0cm]{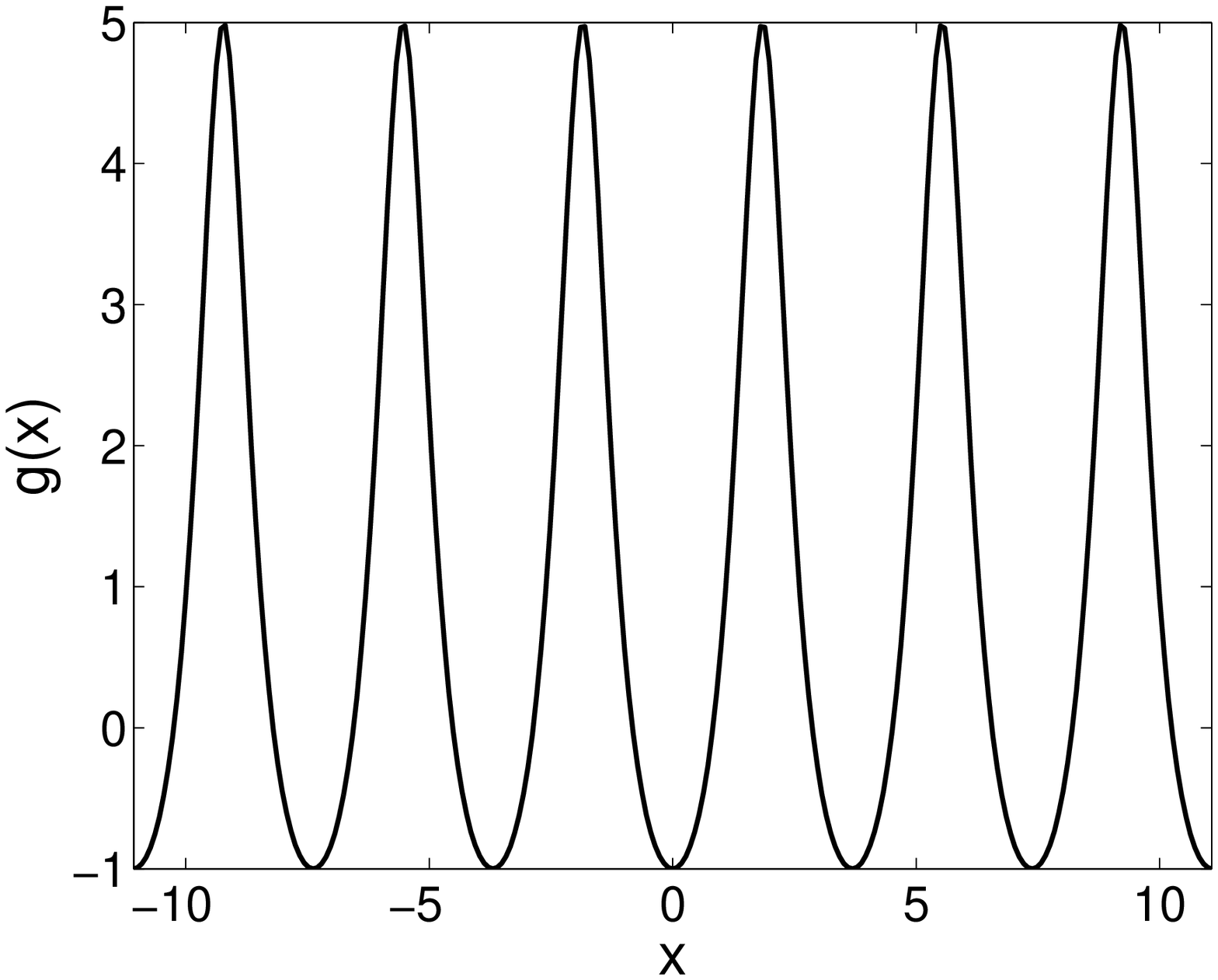}}%
\subfigure[]{\includegraphics [width=5.0cm]{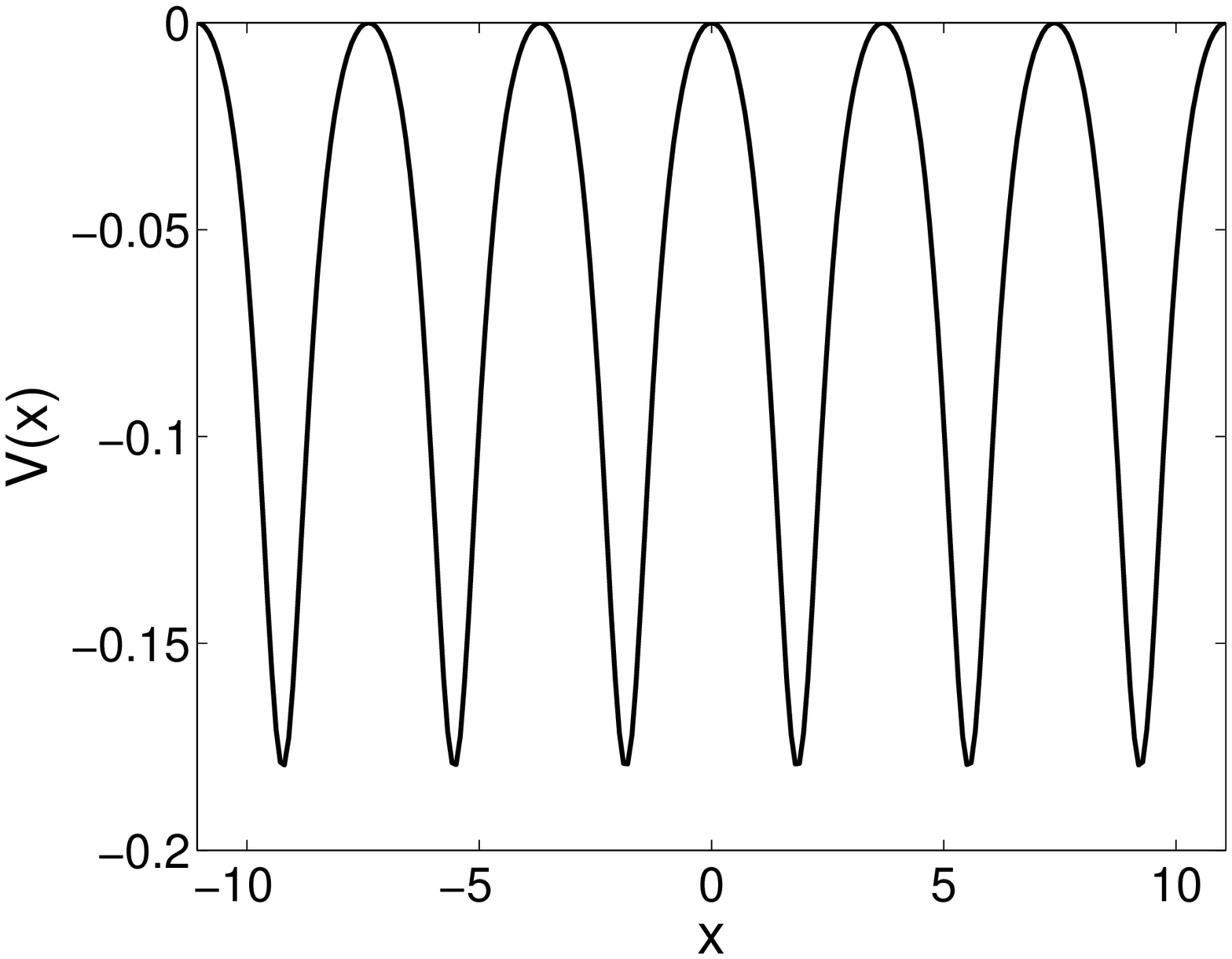}}
\caption{The same as in Figs. \protect\ref{fig6},
\protect\ref{fig7}, and
\protect\ref{fig9}, but for a typical stable solution of the \textrm{%
sn} type found in the presence of the linear potential, for $g_{0}=-1$, $%
b=-0.8$, $r=1$, $g_{1}=2$, and $k=0.7$.}
\label{fig11}
\end{figure}

While, unlike the \textrm{cn} solutions, all patterns of the \textrm{sn}
type tend to be unstable without the linear potential, in the presence of
the potential they may be readily stabilized, in the case of $g_{0}=-1$ and $%
b<0$, which resembles the respective property of the
\textrm{cn}-type patterns, see Fig. \ref{fig4}. An example of the
stable \textrm{sn} structure is shown, for this case, in Fig.
\ref{fig11}. As is typical for other types of stable patterns, in
this figure we observe that maxima of
the density coincide with maxima of $g(x)$ and minima of $V(x)$, cf. Figs. %
\ref{fig6} and \ref{fig7}.

\begin{figure}[tbp]
\center\includegraphics [width=8cm]{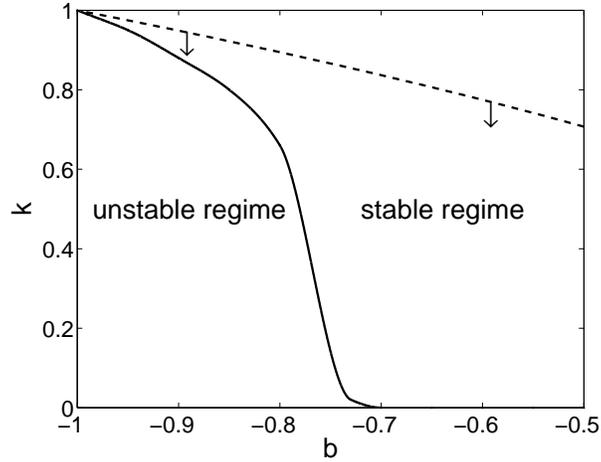} \caption{The
region of the stability for the \textrm{sn}-type periodic
stationary solutions, supported by the linear potential, as found, for $%
g_{0}=-1$ and $g_{1}=2$, in the plane of parameters $b$ and $k$,
which are defined in Eq. (\protect\ref{gsn}), cf. Fig. 4.}
\label{fig12}
\end{figure}

A stability region, summarizing the results of many runs of the simulations,
can also be identified in this case, as shown in Fig. \ref{fig12}, cf. the
stability diagram shown in Fig. \ref{fig4} for the \textrm{cn} solutions
(without the linear potential). The dashed curve with arrows designates the
existence border for the solutions of this type, as determined by condition $%
A^{2}>0$. It follows from Eq. (\ref{AVsn}), that, in the case shown in Fig. %
\ref{fig12}, this existence condition amounts to $k^{2}<|b|$.

\section{Solitons}

\subsection{Stationary solutions for bright solitons}

In the limit of $k\rightarrow 1$, all the above types of the exact periodic
solutions go over into localized structures (solitons) pinned by the
corresponding localized pseudopotential, possibly in the combination with
the localized (trapping) linear potential. In this limit, the solutions of
both the \textrm{cn} and \textrm{dn} types, as given by Eqs. (\ref{solution}%
), (\ref{A}), (\ref{mu}), or (\ref{solutiondn}), (\ref{Adn}), (\ref{mudn}),
reduce to a common expression for the bright soliton,%
\begin{equation}
\psi _{\mathrm{sol}}(x)=\sqrt{\frac{g_{0}(1+2b)}{\cosh ^{2}x+b}},
\label{sol}
\end{equation}%
with $\mu _{\mathrm{sol}}=-1/2$. The nonlinearity-modulation profile which
supports this solution, in the absence of the linear potential, is given by
Eqs. (\ref{g}) and (\ref{g1}), in which $\mathrm{cn}(x)$ is replaced by $%
\mathrm{sech~}x$, i.e.,%
\begin{equation}
g(x)=\frac{g_{0}+g_{1}\mathrm{sech}^{2}x}{1+b\hspace{0.05in}\mathrm{sech}%
^{2}(x)},  \label{gsoliton}
\end{equation}%
and coefficient $g_{1}$ is replaced by its limit form for $k=1$, i.e., as
follows form Eq. (\ref{g1}),
\begin{equation}
g_{1}(k=1)=\frac{g_{0}b}{2}\frac{b-1}{2b+1}.  \label{g1sol}
\end{equation}

The soliton may be stable only if $g(x)$ given by Eq. (\ref{gsoliton}) has a
\emph{maximum} at $x=0$ [for a minimum of $g(x)$ at $x=0$, an obvious
variational consideration of energy (\ref{E}) immediately predicts an
instability of the soliton placed at the maximum of the respective
pseudopotential]. The general condition for the existence of the maximum of $%
g(x)$ at $x=0$ was actually obtained above, $g_{1}>g_{0}b$. In the present
case, with $g_{1}$ taken as per Eq. (\ref{g1sol}), it amounts to $-1<b<0$,
which is, thus, a necessary condition for the stability of the soliton.
Since solution (\ref{sol}) exists only for $g_{0}\left( 1+2b\right) >0$, we
finally conclude that the solitons may be stable only in the case of%
\begin{equation}
g_{0}=-1,~-1/2<b<-1.  \label{sol-stab}
\end{equation}%
In fact, the nonlinearity is sign-changing in this case, because $g(x=\pm
\infty )=g_{0}=-1$, while $g(x=0)=(1/2)(2-|b|)/(2|b|-1)>0$. Lastly, the norm
of soliton (\ref{sol}) is
\begin{equation}
N_{\mathrm{sol}}\equiv \int_{-\infty }^{+\infty }\psi _{\mathrm{sol}%
}^{2}(x)dx=\frac{2\left( 2|b|-1\right) }{\sqrt{|b|\left( 1-|b|\right) }}%
\arctan \left( \sqrt{\frac{|b|}{1-|b|}}\right) .
\end{equation}

A more general bright-soliton solution can be obtained by setting $k=1$ in
the \textrm{cn} or \textrm{dn} solutions including the linear potential,
i.e., Eqs. (\ref{g}), (\ref{solution}), (\ref{V(x)}), (\ref{V0}), (\ref{AV}%
), and (\ref{muV}) [recall that $r$ cannot be set equal to $1$ in this case,
and $g_{1}$ is a free parameter, while Eq. (\ref{g1}) is irrelevant]. This
leads to $\mu =1-r^{2}/2$ and%
\begin{equation}
\psi _{\mathrm{sol}}(x)=r\sqrt{\frac{3\left( 1+b)\right) }{2\left(
g_{0}-g_{1}/b\right) \left[ \cosh ^{2}(rx)+b\right] }}.  \label{sol2}
\end{equation}%
The existence condition for solution (\ref{sol2}) is
\begin{equation}
g_{0}>g_{1}/b,  \label{solexist}
\end{equation}%
which is different from that at which soliton (\ref{sol}) exists.
Further, the nonlinearity-modulation function supporting solution (\ref{sol2}%
) is given by the same expression (\ref{gsoliton}) as above, with the
difference that $g_{1}$ is a free parameter in it, and the respective form
of the linear potential is%
\begin{eqnarray}
V(x) &=&\frac{1+V_{0}\mathrm{sech}^{2}(rx)}{1+b~\mathrm{sech}^{2}(rx)},
\label{Vsoliton} \\
V_{0} &=&b+\frac{r^{2}}{2}\left[ 1-b+3\frac{\left( 1+b\right) g_{1}}{%
g_{0}b-g_{1}}\right] .  \label{V0soliton}
\end{eqnarray}%
Potential (\ref{Vsoliton}) stabilizes the soliton if it represents a
potential well at $x=0$, which means $V_{0}<b$. A straightforward analysis
of expression (\ref{V0soliton}), taking into regard Eq. (\ref{solexist}),
demonstrates that the latter condition is met for%
\begin{equation}
g_{0}\left( b-1\right) >(2/b)\left( 2b+1\right) g_{1}.  \label{well}
\end{equation}

\subsection{The stability of bright solitons}

\begin{figure}[tbp]
\center\subfigure[]{\includegraphics [width=7.0cm]{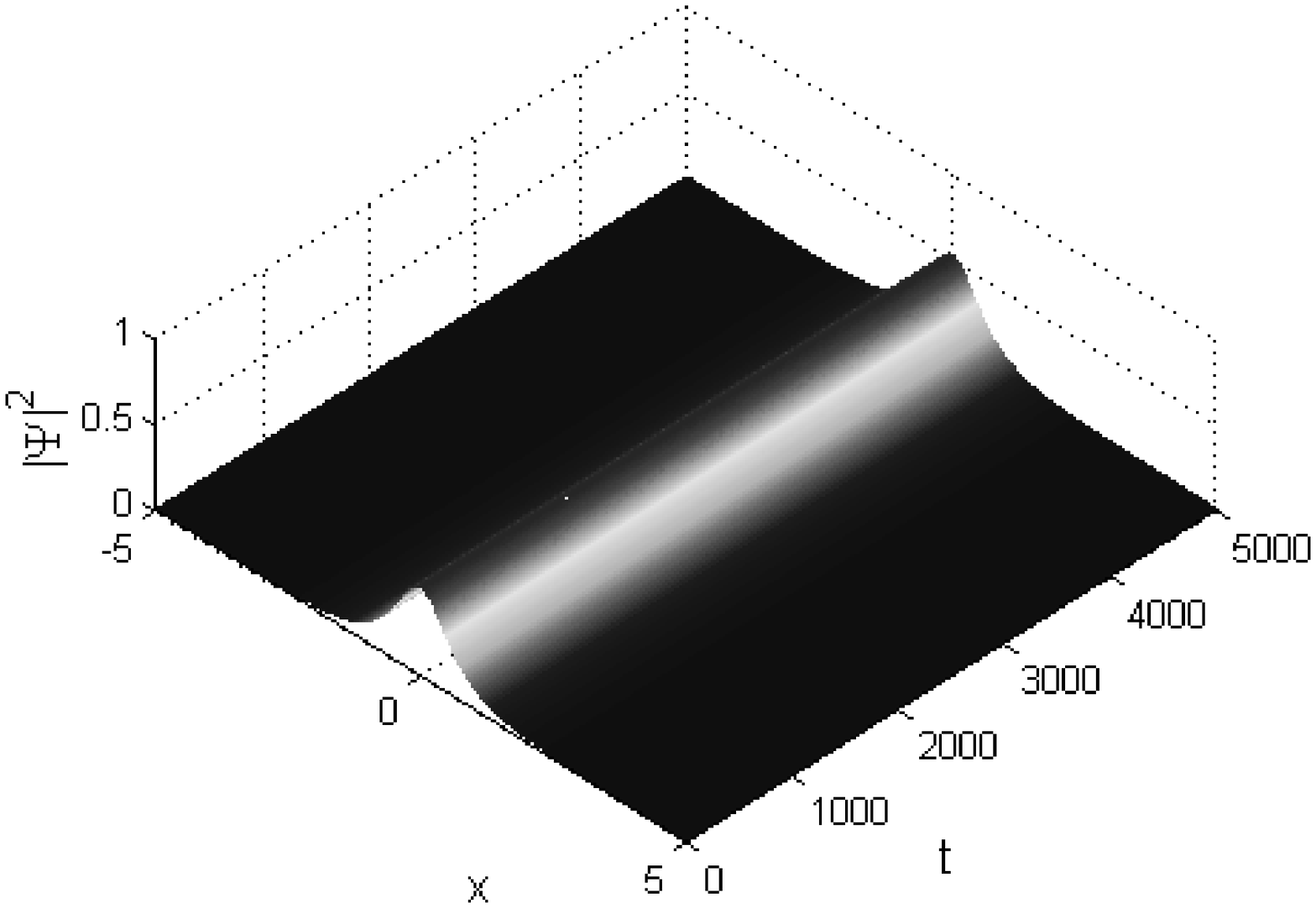}}%
\subfigure[]{\includegraphics [width=5.0cm]{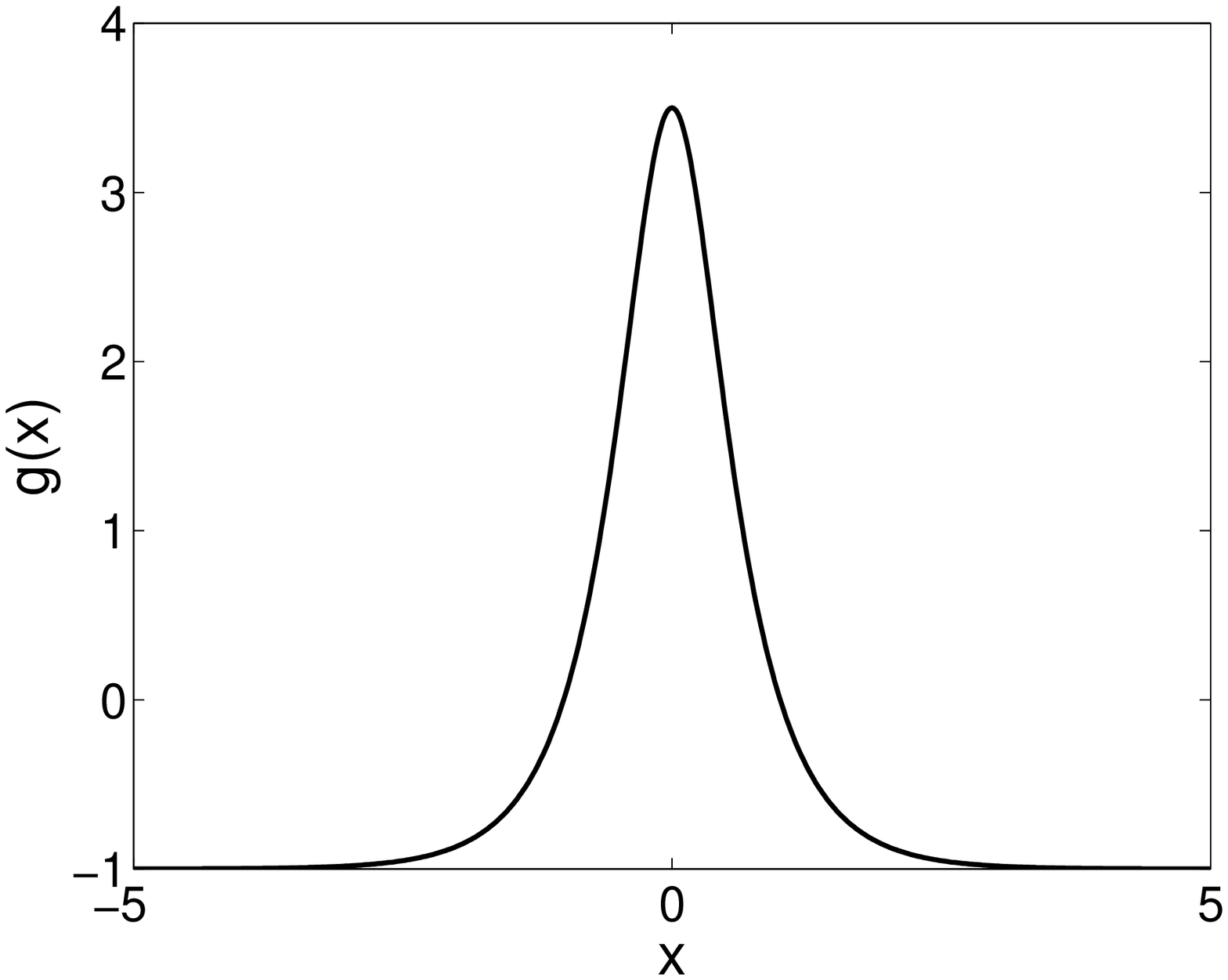}}
\caption{The same as in Fig. \protect\ref{fig2}, but for a case of a \emph{%
stable bright soliton}, obtained as a limit of the \textrm{cn}
solutions with $k=1$ [accordingly, panel (b) displays the localized
{\it sign-changing} nonlinearity-modulation function $g(x)$ which
supports the soliton]. Other parameters are $g_{0}=-1$, $b=-0.6$,
$r=1$, $g_{1}=2.4$.} \label{fig13}
\end{figure}

Systematic simulations of Eq. (\ref{GPE}) with $V(x)=0$, starting from the
initial condition corresponding to a perturbed bright soliton, demonstrate
that the solitons are indeed stable in region (\ref{sol-stab}), where this
should be expected, as argued above. An example of the stable soliton is
shown in Fig. \ref{fig13}. For values of $b$ approaching $-1$, see Eq. (\ref%
{sol-stab}), the initially imposed random perturbations gradually grow,
which, however, is interpreted as the growing sensitivity of the solution to
the random perturbations, rather than as a true instability. A technical
definition of the ``low sensitivity" to the random noise may be adopted in
the form of the requirement that the relative amplitude of the perturbations
must remain below $3\%$, after the evolution time $t=5000$, if the initial
perturbation amplitude was $1\%$, as fixed above. In particular, for the
same parameters as in Fig. \ref{fig13}, $r=1$ and $g_{1}=2.4$, the so
defined sensitivity border is found at $b=-0.75$. For instance, the final
perturbation amplitude is $2.82\%$ and $3.36\%$, at $b=-0.7$ and $-0.8$,
respectively.

\begin{figure}[tbp]
\center\subfigure[]{\includegraphics [width=7.0cm]{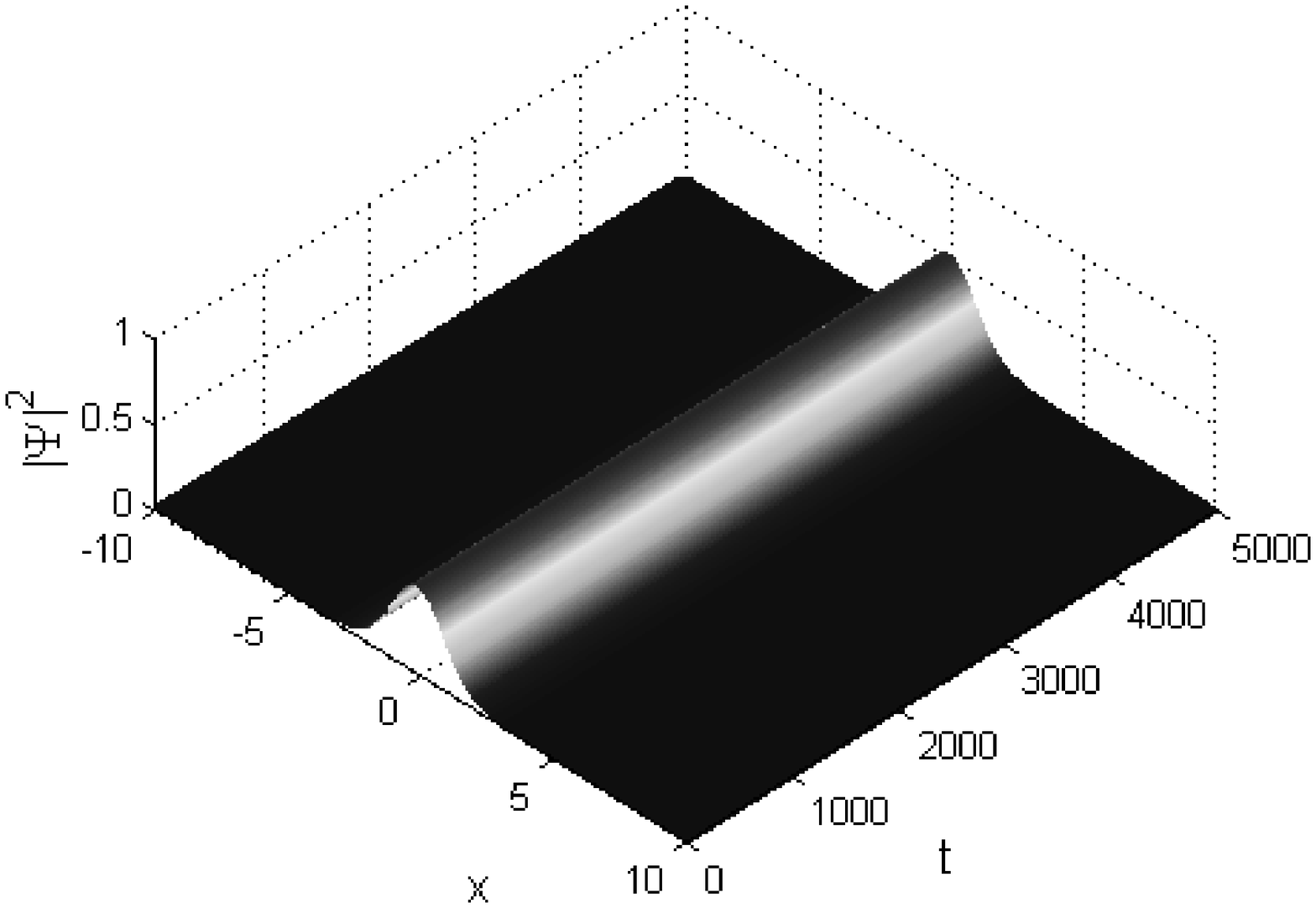}}
\par
\subfigure[]{\includegraphics [width=5.0cm]{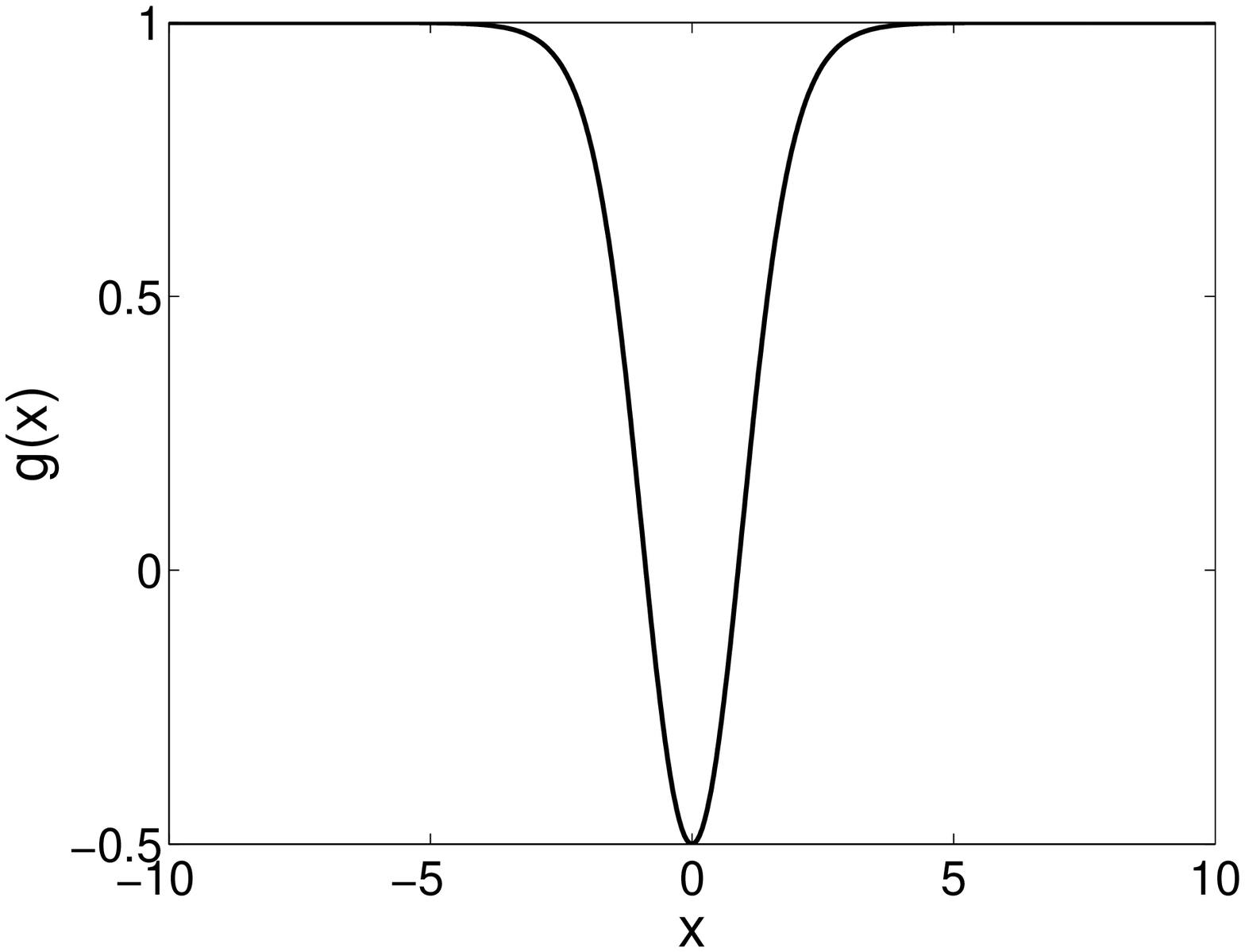}}%
\subfigure[]{\includegraphics [width=5.0cm]{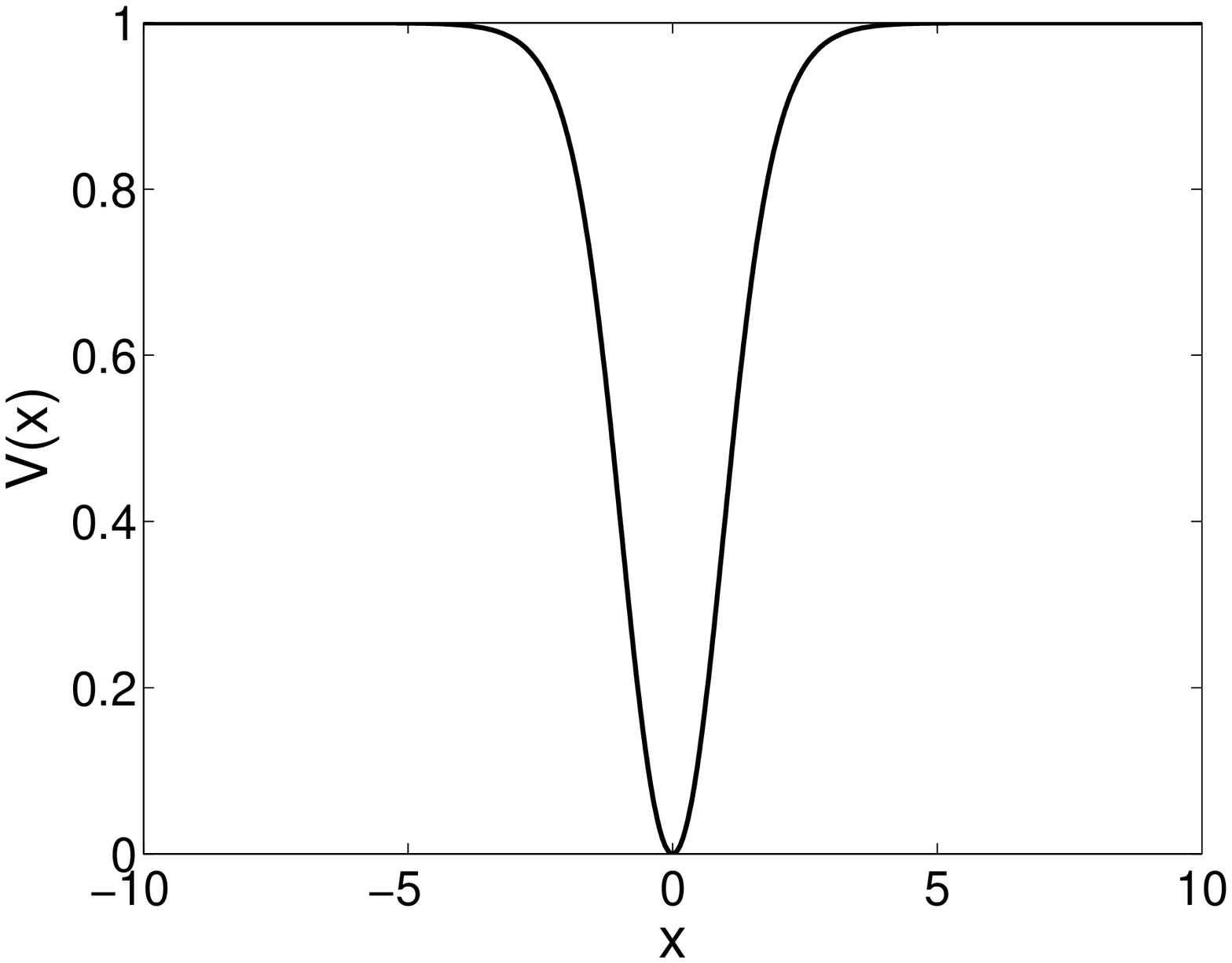}}
\caption{The same as in Figs. \protect\ref{fig6}, \protect\ref{fig7}
and \protect\ref{fig9}, but for a stable soliton found in the
presence of the linear potential. The parameters are $g_{0}=+1$,
$b=1$, $r=1$, $g_{1}=-2$.} \label{fig14}
\end{figure}

\begin{figure}[tbp]
\center\subfigure[]{\includegraphics [width=7.0cm]{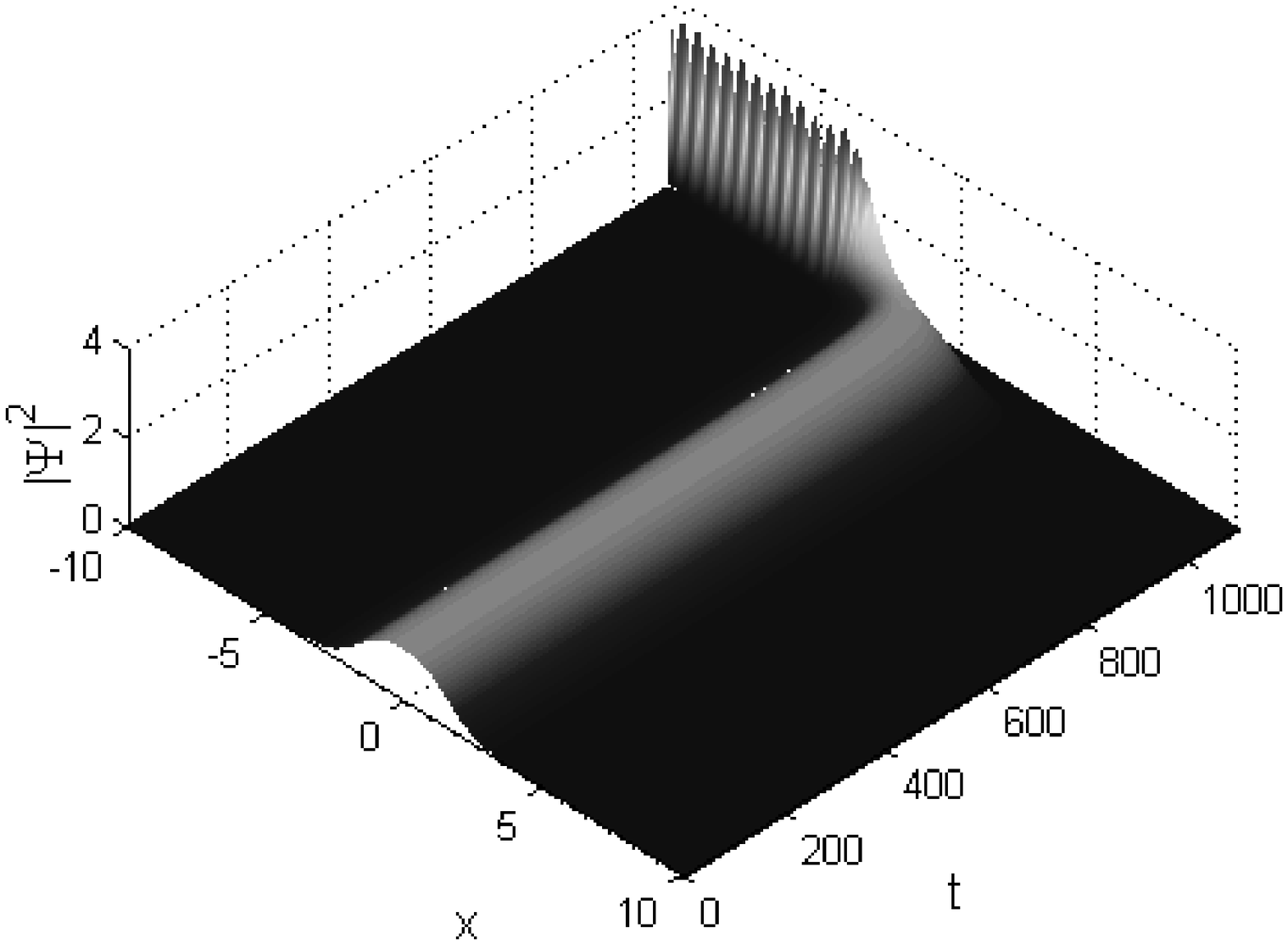}}
\par
\subfigure[]{\includegraphics [width=5.0cm]{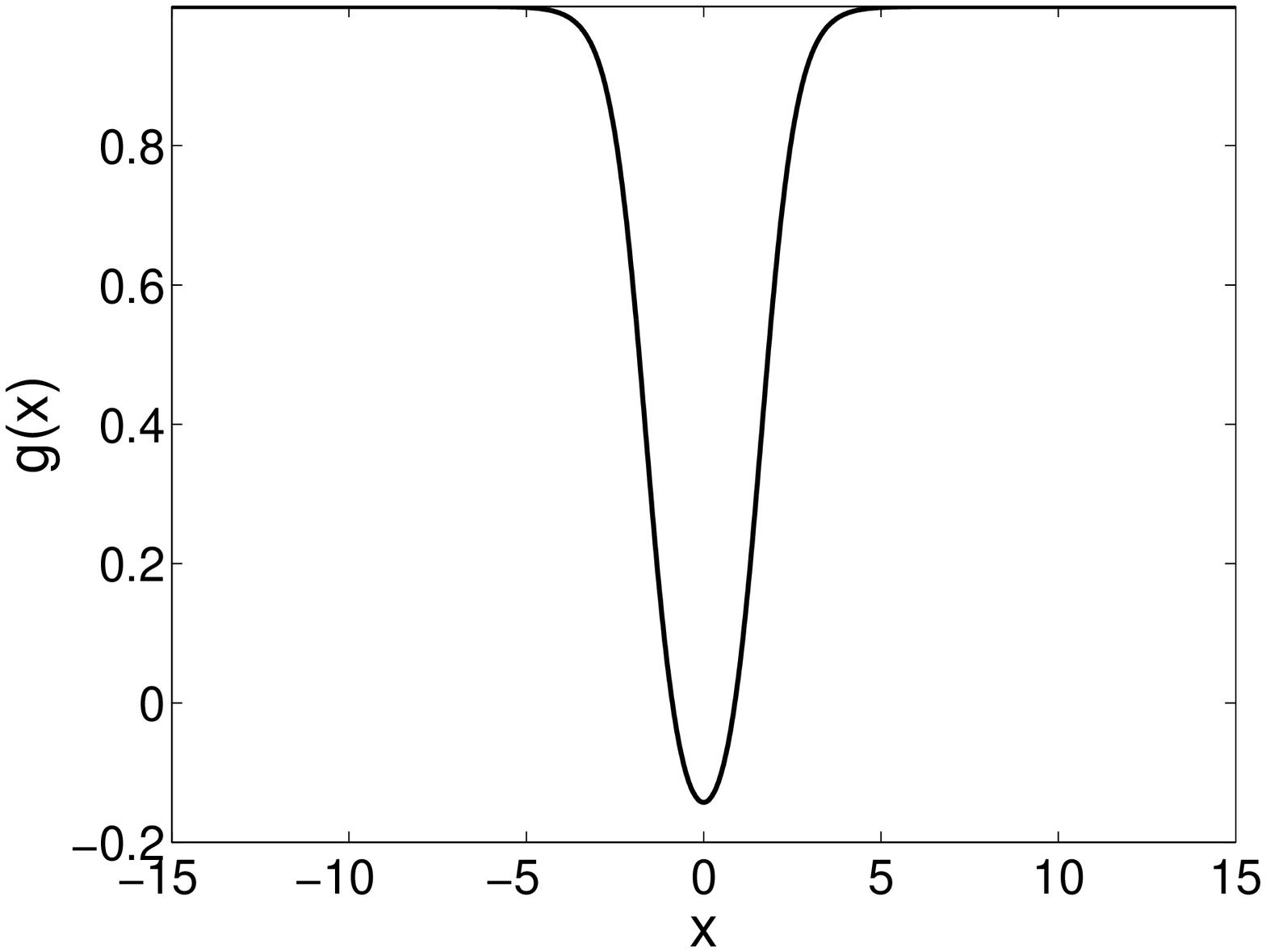}}%
\subfigure[]{\includegraphics [width=5.0cm]{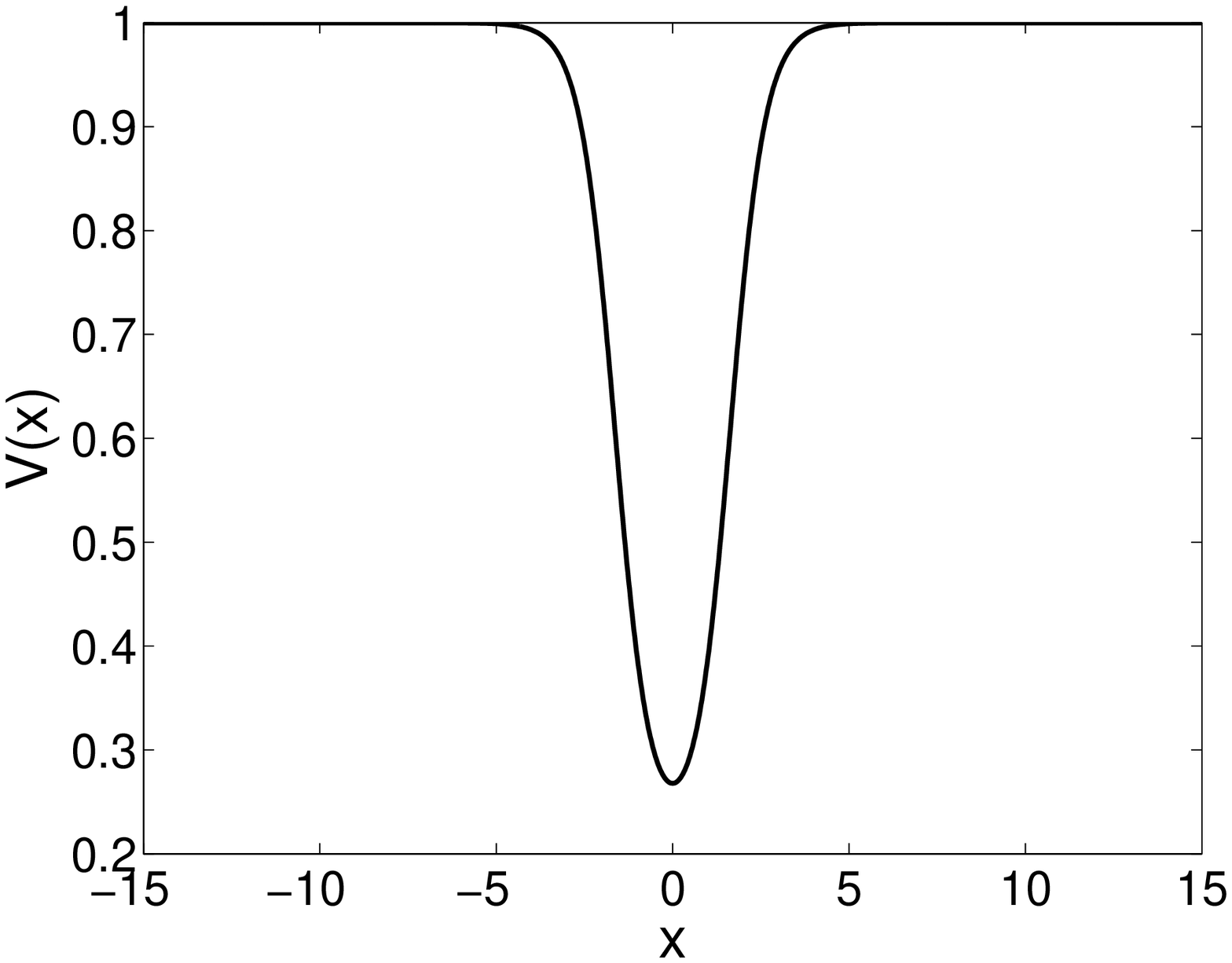}}
\caption{The same as in Fig. \protect\ref{fig14}, but for $b=6$. In
this case, the nonlinear self-repulsive pseudopotential turns out to
be stronger
than the linear trapping potential, hence the bright soliton is \emph{%
unstable}, being spontaneously expelled from the bound state.}
\label{fig15}
\end{figure}

In the presence of the linear potential, stable solitons can be found both
for $g_{0}=-1$, as above, and for $g_{0}=+1$, see Figs. \ref{fig14}-\ref%
{fig16}. For instance, fixing $g_{0}=+1$, $r=1$ and $g_{1}=-2$, as in Figs. %
\ref{fig14} and \ref{fig15}, we find that the solitons are stable for $b<6$,
and unstable for $b\geq 6$ [the soliton is unstable at $b=6$, as shown in
Fig. \ref{fig15}, but still stable--at least, up to $t=5000$ --at $b=5.9$
(not shown here)]. In fact, the transition from the stability to
instability, in the form of a spontaneous escape of the trapped soliton, as
observed in Fig. \ref{fig15}, is explained by the competition between the
self-repulsive localized nonlinear pseudopotential, and the attractive
linear trapping potential, see panels (b) and (c) in Figs. \ref{fig14} and %
\ref{fig15}. Note that parameters corresponding to both figures satisfy
condition (\ref{well}), under which the linear potential is the trapping one.

Lastly, the combination of $g_{0}=-1$ with the linear potential makes both
the nonlinear pseudopotential and linear potential attractive, hence the
solitons are definitely stable in this case, see an example in Fig. \ref%
{fig16}. For the parameters chosen as in this figure, $g_{0}=-1,$ $r=1,$ $%
g_{1}=2$, the above-mentioned technical definition of the sensitivity of the
trapped soliton to the random noise (the perturbation amplitude at the level
of $3\%$ by $t=5000$) sets the ``sensitivity border" at $b=-0.65$, the
solitons being ``hypersensitive" at $-0.65<b<-1 $.

\begin{figure}[tbp]
\center\subfigure[]{\includegraphics [width=7.0cm]{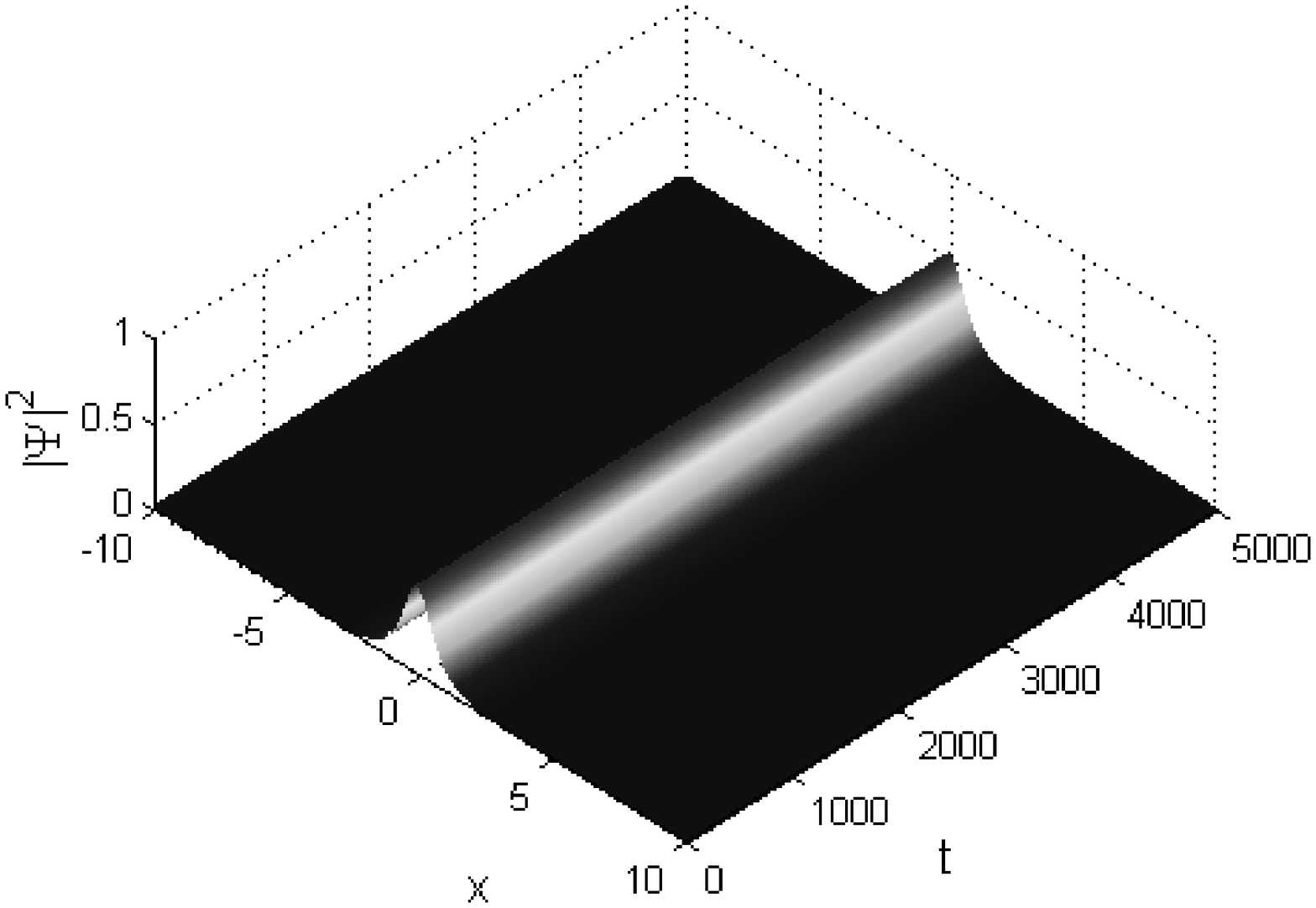}}
\par
\subfigure[]{\includegraphics [width=5.0cm]{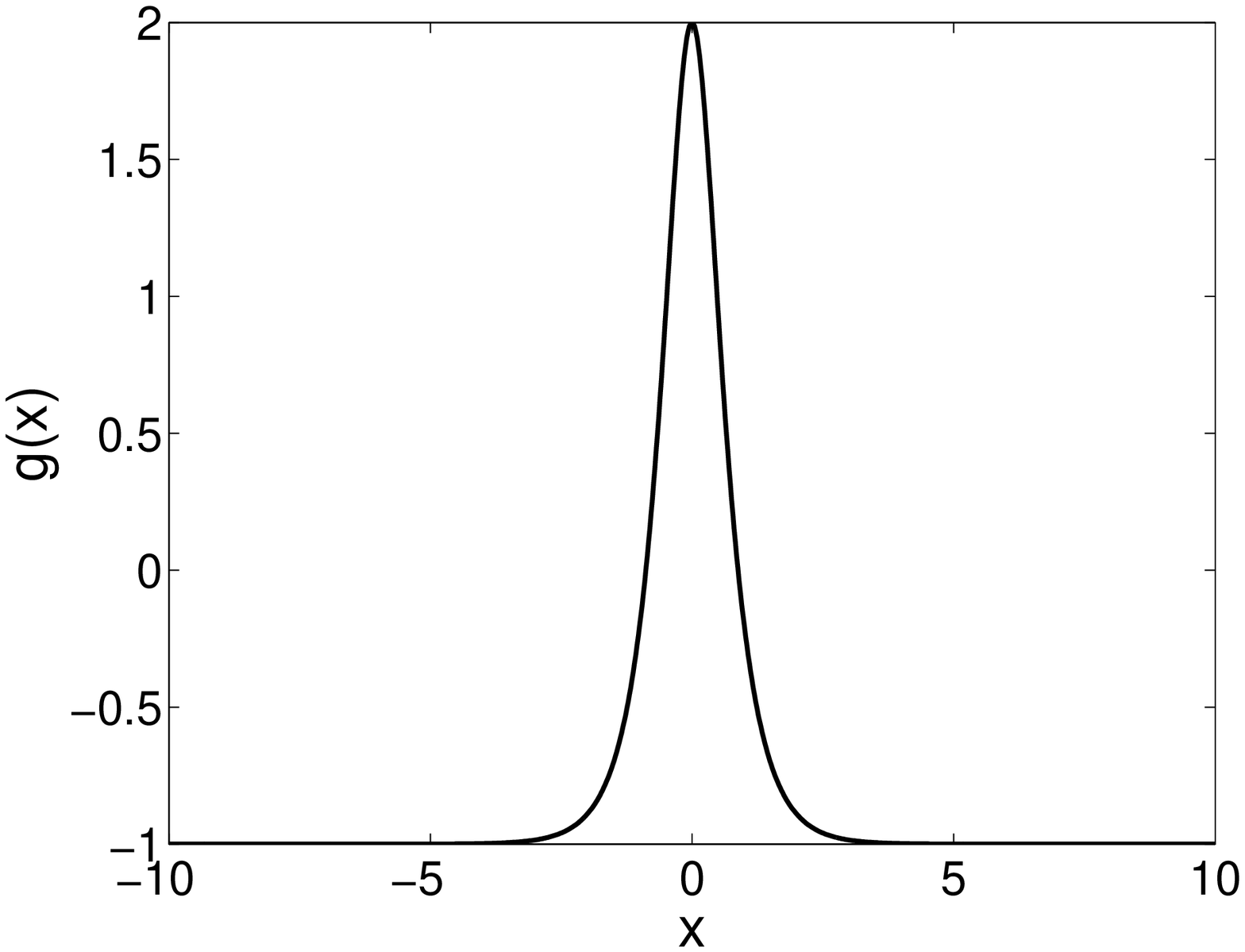}}%
\subfigure[]{\includegraphics [width=5.0cm]{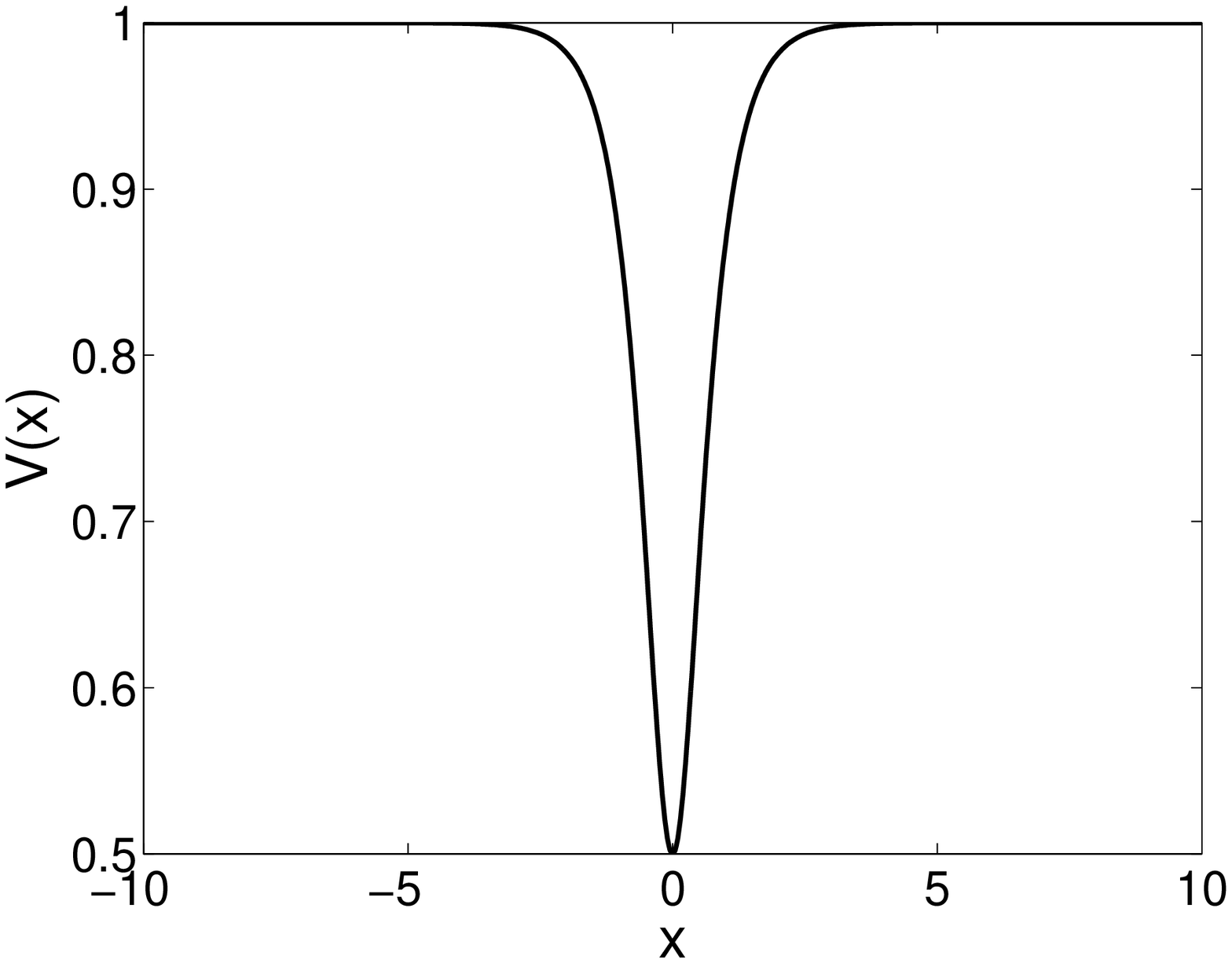}}
\caption{The same as in Fig. \protect\ref{fig14} (also a stable
soliton), but for $g_{0}=-1$, $b=-0.5 $, $r=1$, $g_{1}=2$. Panels
(b) and (c) demonstrate that both the nonlinear pseudopotential and
linear potential are attractive in this case.} \label{fig16}
\end{figure}

\subsection{Dark solitons}

The limit form of the exact solutions of the \textrm{sn} type for $k=1$ are
dark solitons. In the general case, when the respective linear potential (%
\ref{Vsn}) is included, the dark-soliton solution can be obtained from Eqs. (%
\ref{solutionsn}) and (\ref{AVsn}) in the following form:%
\begin{equation}
\psi _{\mathrm{dark-sol}}(x)=\sqrt{\frac{3b}{2\left( g_{1}-g_{0}b\right) }}%
\frac{r(1+b)\tanh \left( rx\right) }{\sqrt{\left( 1+b\right) -b~\mathrm{sech}%
^{2}\left( rx\right) }}.
\end{equation}%
The nonlinearity-modulation function and linear potential which support this
dark soliton are obtained from Eqs. (\ref{gsn}) and (\ref{Vsn}) by setting $%
k=1$ in them:%
\begin{eqnarray}
g(x) &=&\frac{\left( g_{0}+g_{1}\right) -g_{1}\mathrm{sech}^{2}\left(
rx\right) }{\left( 1+b\right) -b~\mathrm{sech}^{2}\left( rx\right) }, \\
V(x) &=&-\frac{\left( U_{0}\right) _{\mathrm{dark-sol}}\tanh ^{2}\left(
rx\right) }{\left( 1+b\right) -b~\mathrm{sech}^{2}\left( rx\right) }, \\
\left( U_{0}\right) _{\mathrm{dark-sol}} &=&\frac{r^{2}}{2}\left[ 1-b\left(
2+3b\right) +\frac{3g_{1}\left( 1+b\right) ^{2}}{\left( g_{0}b-g_{1}\right) }%
\right] .
\end{eqnarray}%
The stability of the dark solitons will be considered elsewhere.

\section{Conclusions}

This work reports three large classes of exact periodic solutions for the
one-dimensional GPE (Gross-Pitaevskii equation) with the pseudopotential
represented by the periodic modulation of the nonlinearity coefficient,
which can be created in the BEC by means of the Feshbach-resonance
technique. In the general case, the model includes a periodic linear-lattice
potential too. The modulation function of the corresponding NL (nonlinear
lattice), the linear potential (if any), and the exact solutions are
expressed in terms of the Jacobi's elliptic functions of three types--%
\textrm{cn}, \textrm{dn}, and \textrm{sn}, which give rise to the three
different classes of the solutions. In the absence of the linear potential,
the solutions depend on two continuous free parameters, $b$ and $k$, and the
sign parameter, $g_{0}$. If the linear potential is included, full solution
families feature two additional continuous parameters, $g_{1}$ and $r$. The
stability of the exact solutions was tested by means of systematic direct
simulations of the perturbed evolution, except for dark solitons, whose
stability will be considered separately.

Subfamilies of the pseudopotentials and respective exact solutions have been
identified with both sign-changing and sign-constant nonlinearity-modulation
functions. Density maxima of the solutions may coincide with minima or
maxima of the periodic pseudopotential; in the former case, the exact
solutions may represent the ground state of the BEC. In the absence of the
linear potential, only the solutions with the density maxima collocated with
minima of the pseudopotential may be stable (although this condition alone
is not sufficient for the stability, in most cases). On the other hand, the
addition of the trapping linear potential may stabilize solutions whose
density maxima coincide with maxima of the nonlinear pseudopotential.

In the limit of the infinite period, the exact solutions reduce to solitons.
The solitons are stable in a large part of parameter regions where they
exist. In particular, they may be stable if located at a maximum of the
nonlinear pseudopotential, provided that the competing trapping linear
potential is strong enough.

This work calls for continuation in other directions. In the models
considered here (with both the periodic and localized pseudopotentials), it
is interesting to find, by means of numerical methods, a full set of
nonlinear states, which should reveal how the exact solutions are embedded
into the full family, and exactly identify the respective ground state.
Further, it is desirable to investigate the dynamical stability of the
solutions in a rigorous form, through the computation of eigenvalues for
modes of small perturbations. Another interesting issue is a possibility of
finding similar families of exact solutions for systems of two or three
coupled GPEs describing multi-component BECs. In fact, the methods used in
earlier works \cite{Kwok2,Kwok21,Kwok22} for obtaining families of exact
solutions in coupled NLSEs can be used in the latter context.

\section*{Acknowledgments}

The work of C.H.T. and K.W.C. is supported by the Research Grants Council of
Hong Kong through contracts HKU 7118/07E and HKU 7120/08E. The work of
B.A.M. was supported, in a part, by grant No. 149/2006 from the
German-Israel Foundation. This author appreciates hospitality of the
Department of Mechanical Engineering at the University of Hong Kong.

\medskip
Received September 2009; revised November 2009.
\medskip

\end{document}